\documentclass[aps,prd,twocolumn,showpacs,nofootinbib,amsmath,amssymb,amsfonts,showkeys]{revtex4}

\usepackage{graphicx}
\usepackage{bm}

\begin{document}
\title{Properties and uncertainties of scalar field models \\ of dark energy with barotropic equation of state}

\author{Bohdan Novosyadlyj}
 \email{novos@astro.franko.lviv.ua}
\author{Olga Sergijenko}
 \email{olka@astro.franko.lviv.ua}
\author{Stepan Apunevych}
 \email{apus@astro.franko.lviv.ua}
\affiliation{Astronomical Observatory of 
Ivan Franko National University of Lviv, Kyryla i Methodia str., 8, Lviv, 79005, Ukraine}
\author{Volodymyr Pelykh}
 \email{pelykh@iapmm.lviv.ua}
\affiliation{Ya. S. Pidstryhach Institute for Applied Problems of Mechanics and Mathematics, \\ Naukova str., 3-b, Lviv, 79060, Ukraine}

\date{\today}

\begin{abstract}
The dynamics of expansion and large scale structure formation in the multicomponent Universe with dark energy modeled by the minimally coupled scalar field with generalized linear barotropic equation of state (EoS) are analyzed. It is shown that the past dynamics of expansion and future of the Universe -- 
eternal accelerated expansion or turnaround and collapse -- are completely defined by the current energy density of a scalar field and relation between its 
current and early EoS parameters. The clustering properties of such models of dark energy and their imprints in the power 
spectrum of matter density perturbations depend on the same relation and, additionally, on the ``effective sound speed'' of a scalar field, defined by its Lagrangian. 
It is concluded that such scalar fields with different values of these parameters are distinguishable in principle. This gives the possibility to constrain 
them by confronting the theoretical predictions with the corresponding observational data. For that we have used the 7-year WMAP  data on CMB anisotropies, the 
Union2 dataset on Supernovae Ia and SDSS DR7 data on luminous red galaxies (LRG) space distribution. Using the Markov Chain Monte Carlo technique the marginalized posterior and mean likelihood distributions are computed for the scalar fields with two different Lagrangians: Klein-Gordon and Dirac-Born-Infeld ones. The properties of such scalar field models of dark energy with best fitting parameters and uncertainties of their determination are also analyzed in the paper. 
\end{abstract}
\pacs{95.36.+x, 98.80.-k}
\keywords{cosmology: dark energy--scalar field--cosmic microwave background--large scale structure of Universe--cosmological parameters}
\maketitle

\section{Introduction}
The unknown nature of dark energy remains the most intriguing problem of physics and cosmology of the last decade. Theorists have proposed a dozen of models well matching the observational data (see e. g. reviews \cite{reviews} and book \cite{textbook}), but key tests for distinguishing of them still have to be developed. This can be achieved by the detailed study of their physical properties and confronting the theoretical predictions of each of them with observations. Physically motivated and historically first after the cosmological constant dark energy candidate is the scalar field violating the strong energy condition. Here we analyze two such fields -- classical and tachyonic -- in order to find out how they match the current observational data assuming that one of these fields dominates our Universe today. As in this paper we are interested mainly in the dark energy effects we restrict ourselves to models with zero curvature of 3-space.
 
Field and fluid approaches are used for inclusion of the scalar field in the cosmological model, which describes the dynamics of expansion of the Universe and formation of its large-scale structure. They are equivalent if the intrinsic entropy of dark energy is taken into account in the equations of perturbations. Here we use the combined approach in which the scalar field as dark energy is described by its dimensionless density, $\Omega_{de}\equiv 8\pi G\rho_{(de)}/3H_0^2$, the equation of state (EoS) parameter $w\equiv P_{(de)}/\rho_{(de)}$ and the effective sound speed\footnote{The terms ``effective sound speed'' and ``adiabatic sound speed'' of dark energy are used in the literature for designations of dark energy intrinsic values which formally correspond to thermodynamical ones.} $c_s^2\equiv\delta P_{(de)}/\delta\rho_{(de)}$, determined by its Lagrangian. We assume also that a scalar field is minimally coupled to other components, i. e. it interacts with them only gravitationally. Therefore these quantities completely define its model. In general case the last two values, $w$ and $c^2_s$, are arbitrary functions of time, which should be defined by additional assumptions about the properties of dark energy. The simplest case $w=const$ and $c_s^2=1$ (classical scalar field) is well studied. The recent determination of value of $w$ for this case by \cite{wmap7}, based on the seven-year WMAP data (WMAP7), the distance measurements from the baryon acoustic oscillations (BAO) in the space distribution of galaxies from SDSS DR7 data \cite{Percival2009}, Hubble constant measurements \cite{Riess2009} and supernova data \cite{SNIa} (WMAP7+BAO+SN) gives the value close to -1: $w=-0.98\pm0.053$ (flat Universe, 68\% CL). So, the cosmological model with such scalar field is rather similar to $\Lambda$CDM-model and has the same problems, e. g. fine tuning and cosmic coincidence. 

Varying in time EoS parameter $w(a)$ of the so-called dynamical dark energy is the more complicated case which can in principle resolve those problems as well as promises the interesting properties of dark energy and its effect on the dynamics and structure formation of the Universe. The dynamical classical scalar fields called ``quintessence'' with different potentials were investigated extensively during for over than a decade (for some early works see \cite{sf}). Often the dependence of EoS parameter $w$ on the scale factor $a$ is assumed in \textit{ad hoc} manner. The linear form of this dependence $w(a)=w_0+(1-a)w_a$ was proposed in \cite{wa} and widely used. Here, $w_0$ and $w_a$ denote the present values of $w$ and its first derivative respectively. The determination of them by \cite{Komatsu2010} on the base of WMAP7+BAO+SN data gives $w_0=-0.93\pm0.12$,  $w_a=-0.41\pm0.72$. Such dark energy evolves from phantom field with $w=-1.34$ at the early epoch to the quintessential one at the current epoch with $w=-0.93$. The dark energy density increases from zero to $\sim1.25\rho^0_{de}$ at $a\approx 0.83$ and decreases asymptotically to zero after that. The previous determination of these parameters by \cite{Komatsu2009} using similar but older datasets gave $w_0=-1.09\pm0.12$, $w_a=0.52\pm0.46$. It means that in this case the dark energy evolves from quintessential field with $w=-0.57$ in the early epoch to the phantom one at current epoch with $w=-1.09$. Its density decreases at early epoch, achieves the minimal value of $\sim0.83\rho^0_{de}$ at the same scale factor $a\approx 0.83$ and grows later. Recently this form has been modified by \cite{Komatsu2009} in order to bring the behavior of dynamical dark energy at early epoch closer to that of $\Lambda$-term. Their approximation has the additional third parameter $a_{tran}$ which, however, is weakly constrained by observations. The mentioned parametrizations of time dependence of EoS parameter allow the phantom divide crossing ($w=-1$) and extend the variety of properties of dark energy and its possible physical interpretations. Other three, four and more parametric approximations of EoS have been analyzed \cite{approx} too. The additional degeneracies and uncertainties of parameters related to the early dark energy density and time variations of EoS parameter are inherent for them. And vice versa, the value of EoS $w$ as well as of energy density $\Omega_{de}$ related to the late epoch are determined well as a result of their main impact on the expansion history of the Universe, horizon scale, distance to CMB last scattering surface and scale-independent growth factor of linear matter density perturbations. These values, however, give no possibility to constrain essentially the types of cosmological scalar fields, or, in other words, the forms of their Lagrangians and potentials.

The additional constraints on the type of cosmological scalar fields can be obtained by determination of the third mentioned above parameter -- the effective  sound speed $c_s^2$, since it is related to the field Lagrangian as $c_s^2=\frac{\partial L}{\partial X}/\left(2X\frac{\partial^2L}{\partial X^2}+\frac{\partial L}{\partial X}\right)$. It governs the evolution of dark energy density perturbations which influence the evolution of matter density ones via the gravitational potential. Its impact on the linear power spectrum of matter density perturbations is essentially lower than the growth factor caused by the background dynamics, but it is scale-dependent and can be appreciable for some types of scalar fields. Some of them are studied carefully in \cite{Putter2010}.

In the previous papers \cite{Novosyadlyj2009,Sergijenko2009b} we have analyzed the cosmological models with minimally coupled classical and tachyonic scalar fields with the EoS parameter varying in time as $w(a)=w_0a^3/(1+w_0-w_0a^3)$ which was deduced from the condition of ``zero adiabatic sound speed'': $c_a^2\equiv\dot P_{(de)}/\dot\rho_{(de)}=0$. It was shown that at the early epoch such fields can mimic the dark matter and diminish the fine tuning problem. The magnitude of dark energy perturbations of such fields increases before entering the acoustic horizon and decays in oscillating manner after that. It was shown also that they affect the magnitude of gravitational potential and cause the scale dependence of matter density perturbations which can be used for constraining the parameters of such dark energy models along with other cosmological ones. 

Here we analyze more general case -- ``constant adiabatic sound speed'' $c_a^2=const$, that is equivalent to assumption of the linear barotropic EoS of dark energy $P_{(de)}=C_1\rho_{(de)}+C_2$, where $C_1$ and $C_2$ are constants. The important question related to such scalar fields is: how their introduction agrees with the observational data on accelerated expansion of the Universe, its large scale structure and CMB anisotropy? To find the answer we perform analysis using the Markov Chain Monte Carlo (MCMC) method for multicomponent cosmological model and the present data, namely seven-year WMAP data on CMB anisotropy, SDSS DR7 and the SNIa photometric relations. 

The paper is organized as follows. In Section \ref{bckgr} we discuss the background dynamics of our models and parametrization of EoS parameter by the ``adiabatic sound speed''. In Section \ref{evolveeq} the evolution equations for the scalar field perturbations in the synchronous gauge are presented. In Section \ref{method_data} the method of determination of cosmological parameters and used observational datasets are described. In Section \ref{results} we present the results of determination of scalar field parameters and discuss the properties of models with the best fitting ones. The conclusion can be found in Section \ref{conclusions}.  

\section{Dark energy with barotropic EoS and background equations}\label{bckgr}
We assume that background Universe is spatially flat, homogeneous and isotropic with
Friedmann-Robertson-Walker (FRW) metric of 4-space
$$ds^2=g_{ij} dx^i dx^j =a^2(\eta)(d\eta^2-\delta_{\alpha\beta} dx^{\alpha}dx^{\beta}),$$
where $\eta$ is the conformal time defined by relation\footnote{Here and below we put $c=1$.}
$dt=a(\eta)d\eta$ and
$a(\eta)$ is the scale factor, normalized to 1 at the current
epoch $\eta_0$. The Latin indices $i,\,j,\,...$ run from 0 to 3 and
the Greek ones are used for the spatial part of the metric: $\alpha,\,
\beta,\,...$=1, 2, 3. 
\begin{figure}
\includegraphics[height=.5\textheight]{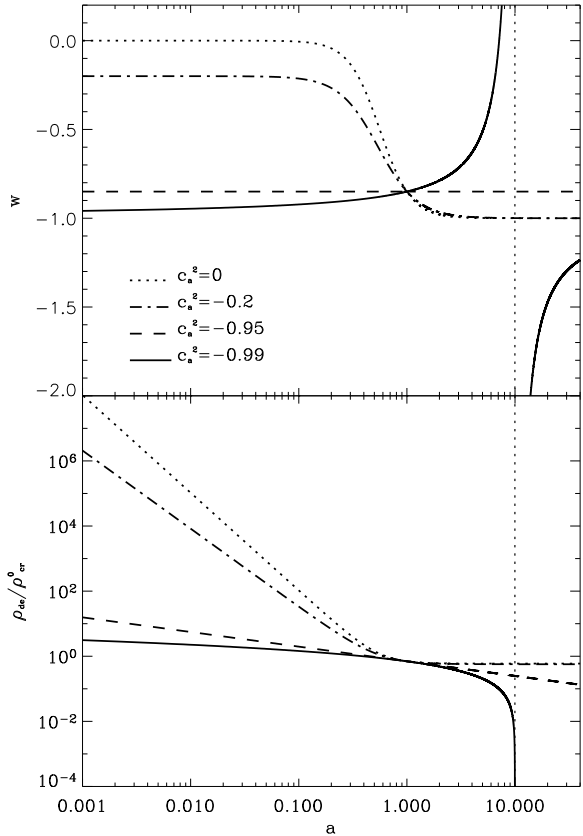}
\caption{Top panel: the dependences of EoS parameter on scale factor for barotropic dark energy with $w_0$=-0.85 and different $c_a^2$ (0, -0.2, -0.85, -0.99).  Bottom panel: the dependences of dark energy density (in the units of critical one at the current epoch) on scale factor with EoS parameters presented in the top panel.}
\label{fig1}
\end{figure}
\begin{figure}
\includegraphics[height=.5\textheight]{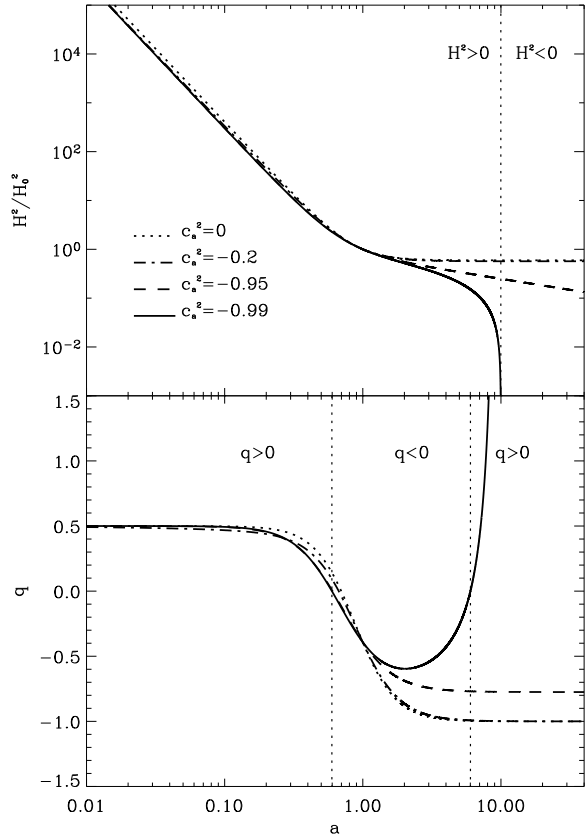}
\caption{The dynamics of expansion of the Universe with barotropic dark energy:  $H^2(a)$ (top panel) and $q(a)$ (bottom one) for the same models as in Fig. \ref{fig1}.}
\label{fig2}
\end{figure}
We also suppose that the Universe is filled with non-relativistic particles (cold dark matter and baryons), relativistic ones (thermal electromagnetic radiation and massless neutrino) and minimally coupled dark energy. The last one is considered as the scalar field with either Klein-Gordon (classical) or Dirac-Born-Infeld (tachyonic) Lagrangian 
\begin{eqnarray}
L_{clas}=X-U(\phi),\,\,\,
L_{tach}=-\tilde{U}(\xi)\sqrt{1-2\tilde{X}},\label{L}
\end{eqnarray}
where ${U}(\phi)$ and $\tilde{U}(\xi)$ are the field potentials defining the model of the scalar field, $X=\phi_{;i}\phi^{;i}/2$ and $\tilde{X}=\xi_{;i}\xi^{;i}/2$ are kinetic terms. We assume also that the background scalar fields are homogeneous ($\phi(\textbf{x},\eta)=\phi(\eta)$, $\xi(\textbf{x},\eta)=\xi(\eta)$), so their energy density and pressure depend only on time:
\begin{eqnarray}
&&\rho_{clas}=X+{U}(\phi),\,\,\,\,\,
P_{clas}=X-{U}(\phi),\\
&&\rho_{tach}=\frac{\tilde{U}(\xi)}{\sqrt{1-2\tilde{X}}},\,\,\,\,
P_{tach}=-\tilde{U}(\xi)\sqrt{1-2\tilde{X}}.
\end{eqnarray}
The EoS parameters $w_{de}\equiv P_{de}/\rho_{de}$ for these fields are following:
\begin{eqnarray}
 w_{clas}=\frac{X-U}{X+U},\,\,\,\,\,w_{tach}=2\tilde{X}-1\label{w_X}.
\end{eqnarray}
Using the last relations the field variables and potentials can be presented in terms of densities and EoS parameters as:
\begin{eqnarray}
&&\phi(a)-\phi_0=\pm\int_1^a\frac{da'\sqrt{\rho_{de}(a')(1+w(a'))}}{a'H(a')},\label{vcl}\\
&&U(a)=\frac{\rho_{de}(a)\left[1-w(a)\right]}{2}\label{pcl}
\end{eqnarray}
for the classical Lagrangian and
\begin{eqnarray}
&&\xi(a)-\xi_0=\pm\int_1^a\frac{da'\sqrt{1+w(a')}}{a'H(a')},\label{vt}\\
&&\tilde{U}(a)=\rho_{de}(a)\sqrt{-w(a)}\label{pt}
\end{eqnarray}
for the tachyonic one.

The dynamics of expansion of the Universe is completely described by the Einstein equations
\begin{eqnarray}
R_{ij}-{\frac{1}{2}}g_{ij}R=8\pi G \left(T_{ij}^{(m)}+T_{ij}^{(r)}+T_{ij}^{(de)}\right),
\label{E_eq}
\end{eqnarray}
where $R_{ij}$ is the Ricci tensor and $T_{ij}^{(m)}$, $T_{ij}^{(r)}$, $T_{ij}^{(de)}$ are the energy-momentum tensors of non-relativistic matter $(m)$, relativistic matter $(r)$, and dark energy, $(de)$, correspondingly. Assuming that the interaction between these components is only gravitational, each of them should satisfy the differential energy-momentum conservation law separately:
\begin{eqnarray}
T^{i\;\;(n)}_{j\;;i}=0.
\end{eqnarray}
Hereafter ``;'' denotes the covariant derivative with respect to the
coordinate with given index in the space with metric $g_{ij}$ and $(n)$
stands for $(m)$, $(r)$ or $(de)$.

For the perfect fluid with density $\rho_{(n)}$ and pressure $P_{(n)}$, related by the equation of state
$P_{(n)}=w_{(n)}\rho_{(n)}$, it gives
\begin{eqnarray}
\dot{\rho}_{(n)}=-3\frac{\dot a}{a} \rho_{(n)}(1+w_{(n)})\label{rho'},
\end{eqnarray}
here and below a dot denotes the derivative with respect to the conformal time, ``$\dot{\;\;}$''$\equiv d/d\eta$. For the non-relativistic matter $w_{(m)}=0$ and $\rho_{(m)}=\rho_{(m)}^{(0)}a^{-3}$, for the relativistic one $w_{(r)}=1/3$ and $\rho_{(r)}=\rho_{(r)}^{(0)}a^{-4}$. Hereafter ``0'' denotes the current values.
\begin{figure*}
\includegraphics[width=.47\textwidth]{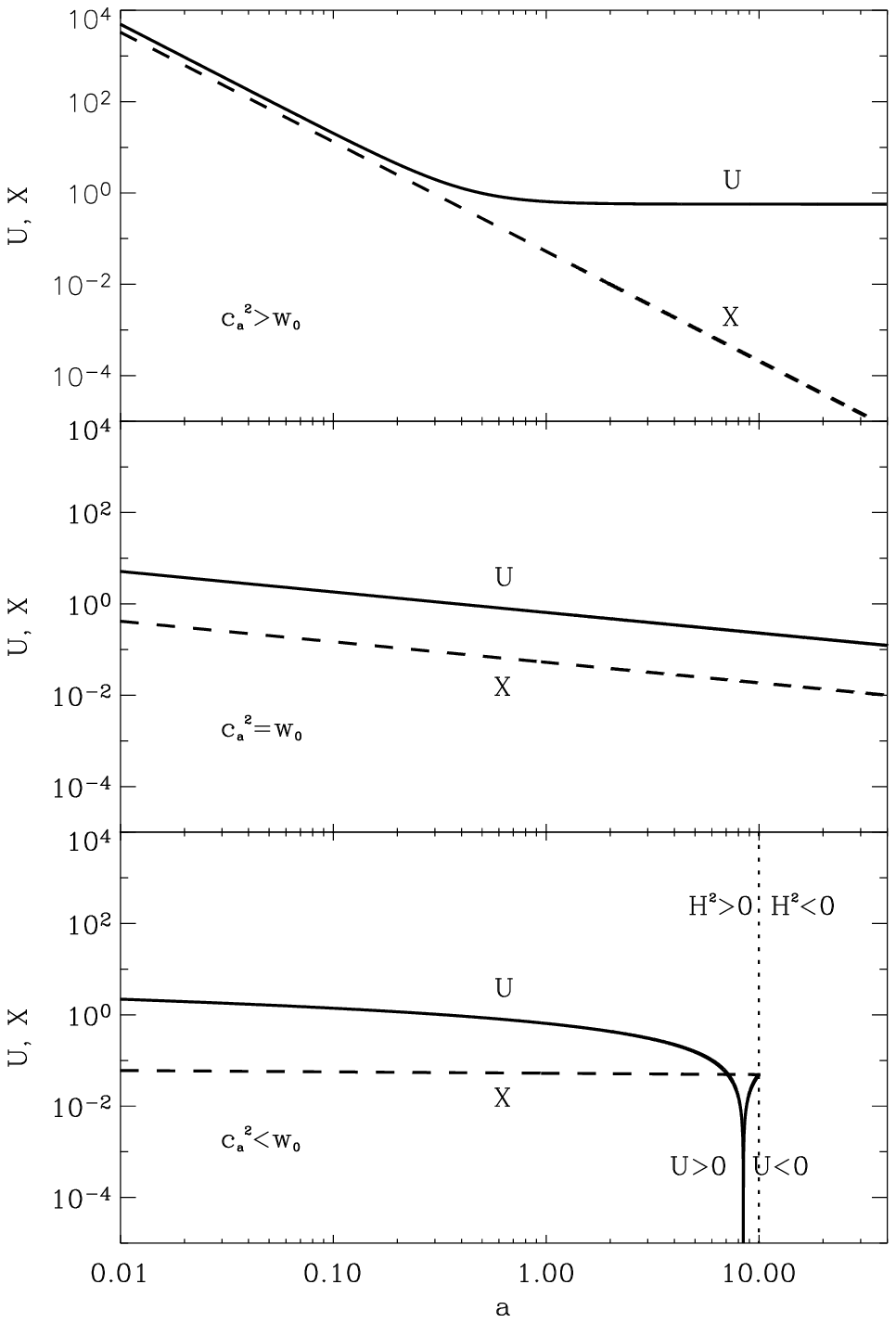}
\includegraphics[width=.47\textwidth]{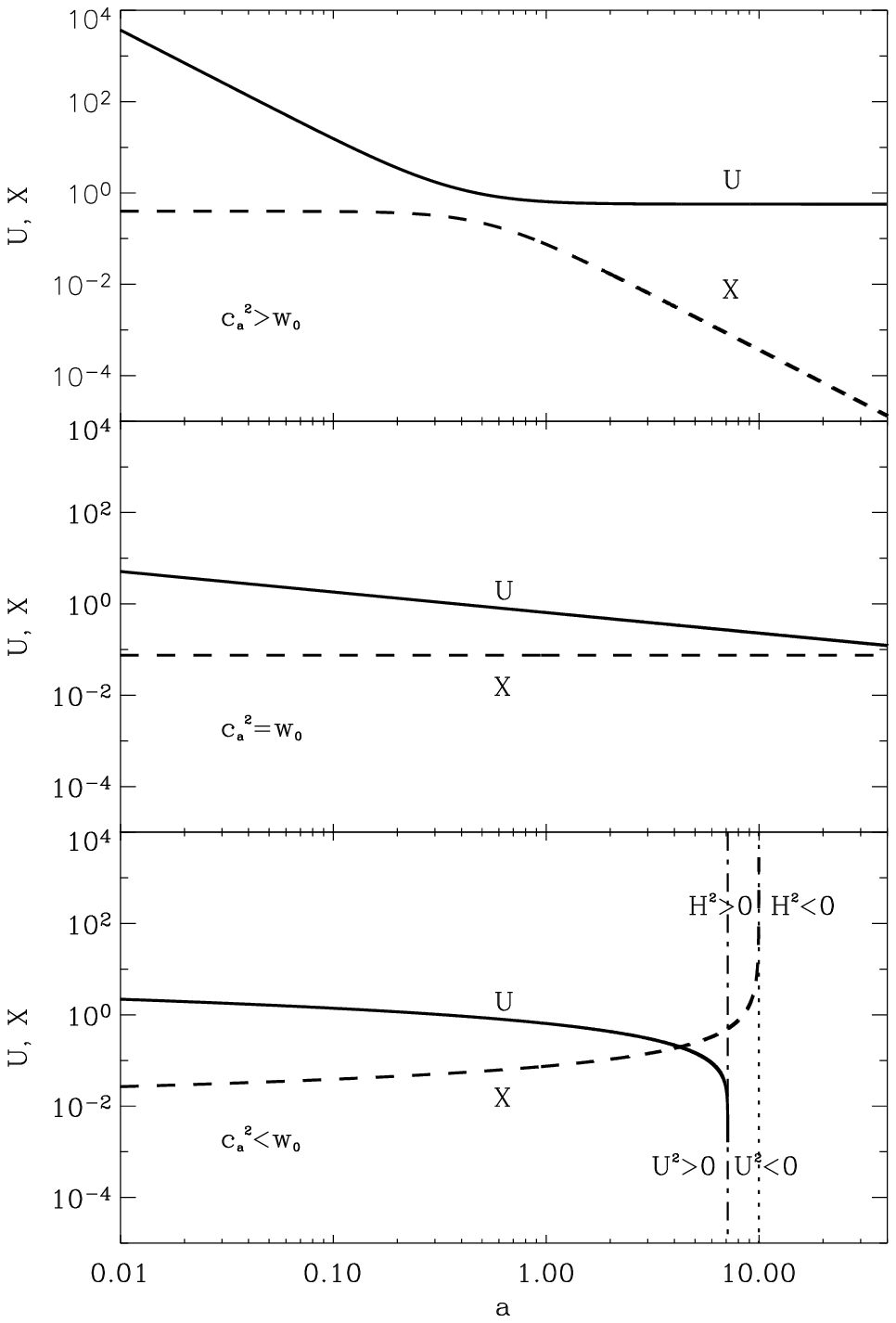}
\caption{Evolution of potentials and kinetic terms of classical (left) and tachyonic (right) scalar fields with barotropic EoS for $c_a^2>w_0$ (top panels), $c_a^2=w_0$ (middle panels) and $c_a^2>w_0$ (bottom panels). In the top panels $c_a^2=-0.2$, in the middle ones $c_a^2=-0.85$ and in the bottom ones $c_a^2=-0.99$. In all panels $w_0=-0.85$.}
\label{U_X}
\end{figure*}
The EoS parameter $w_{(de)}$ and adiabatic sound speed $c_{a\;(de)}^2\equiv\dot{P}_{(de)}/\dot{\rho}_{(de)}$ are related by the ordinary differential equation:
\begin{eqnarray}
 w'=3a^{-1}(1+w)(w-c_a^2),\label{w'}
\end{eqnarray}
where a prime denotes the derivative with respect to the scale factor $a$.  Here and below we omit index $de$ for $w_{(de)}$ and $c_{a\;(de)}^2$. As it can be easily seen, the derivative of EoS parameter with respect to the scale factor will be negative for $w<c_a^2$ and positive for $w>c_a^2$. In the first case the repulsive properties of scalar fields will be raising, in the second one -- receding. In the general case $c_a^2$ can be an arbitrary function of time, but here we assume that it is constant: $c_a^2=const$. In such case the temporal derivative of $P_{(de)}(\eta)$ is proportional to the temporal derivative of $\rho_{(de)}(\eta)$. The integral form of this condition is the generalized linear barotropic equation of state
\begin{eqnarray}
 P_{(de)}=c_a^2\rho_{(de)}+C,\label{beos}
\end{eqnarray} 
where $C$ is a constant. Cosmological scenarios for the Universe filled with the fluid with such EoS\footnote{Often called in literature ``wet dark fluid''.} have been
analyzed in \cite{Babichev2005}. The solution of the differential equation (\ref{w'}) for $c_a^2=const$ is following:
\begin{eqnarray}
 w(a)=\frac{(1+c^2_a)(1+w_0)}{1+w_0-(w_0-c^2_a)a^{3(1+c^2_a)}}-1,\label{w}
\end{eqnarray}
where the integration constant of (\ref{w'}) $w_0$ is chosen as the current value of $w$. One can easily find that (\ref{w}) gives (\ref{beos}) with $C=\rho_{(de)}^{(0)}(w_0-c_a^2)$, where $\rho_{(de)}^{(0)}$ is current density of dark energy.
Thus, we have two values $w_0$ and $c_a^2$ defining the EoS parameter $w$ at any redshift $z=a^{-1}-1$. The time dependences of barotropic EoS parameter  for different values of $c_a^2$ are shown in the top panel of Fig. \ref{fig1}. As it follows from (\ref{w}),  $c_a^2$ corresponds to the EoS parameter at the beginning of expansion ($w_{init}\equiv w(0)=c_a^2,\,a=0,\,z=\infty$).  

The differential equation (\ref{rho'}) with $w$ from (\ref{w}) has the analytic solution too:
\begin{eqnarray}
\rho_{(de)}=\rho_{(de)}^{(0)}\frac{(1+w_0)a^{-3(1+c_a^2)}+c_a^2-w_0}{1+c_a^2}.\label{rho}
\end{eqnarray}
The dependences of dark energy density on scale factor for different values of $c_a^2$ 
are shown in the bottom panel of Fig. \ref{fig1}. 
\begin{figure*}
\includegraphics[width=.47\textwidth]{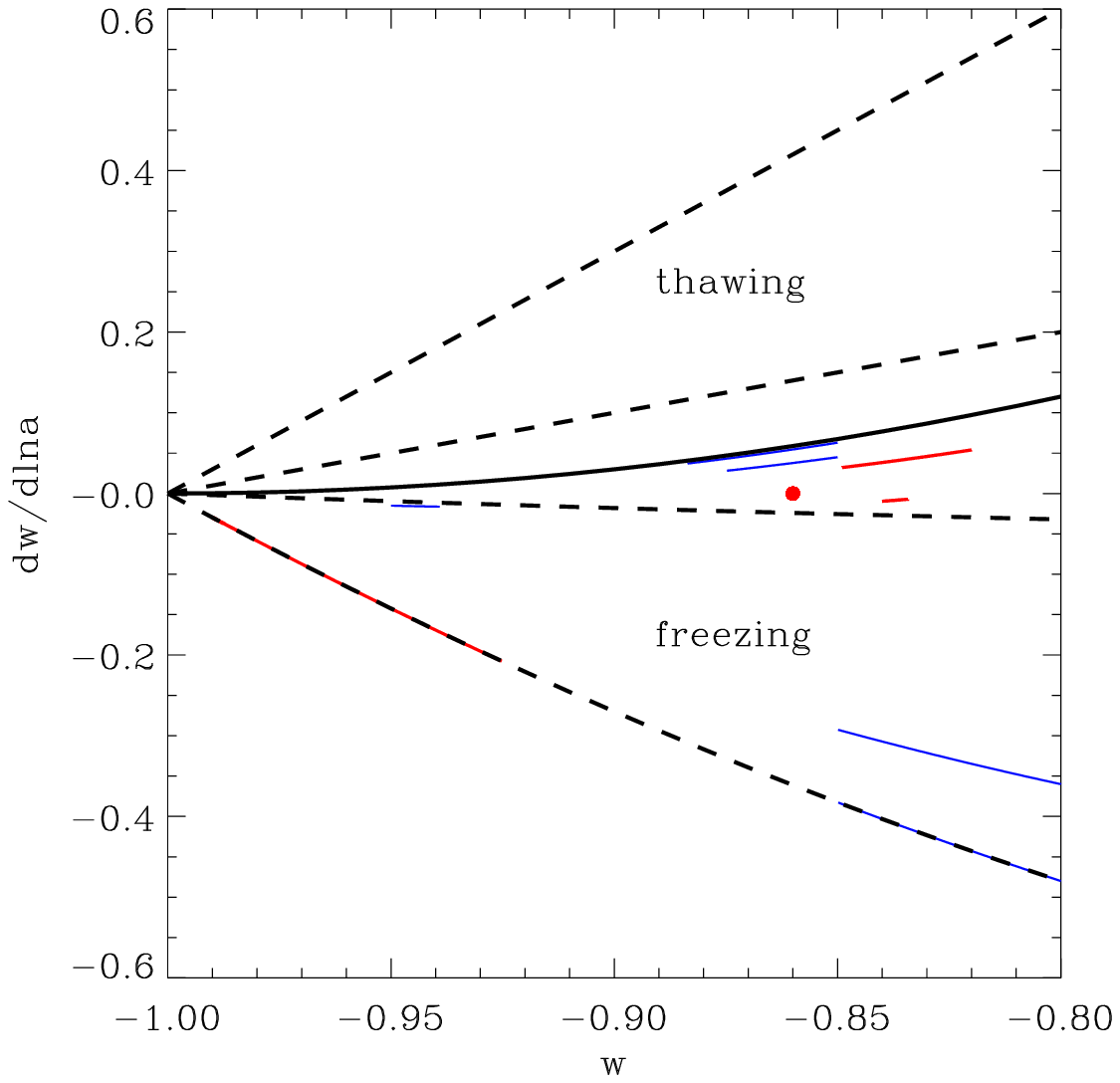}
\includegraphics[width=.47\textwidth]{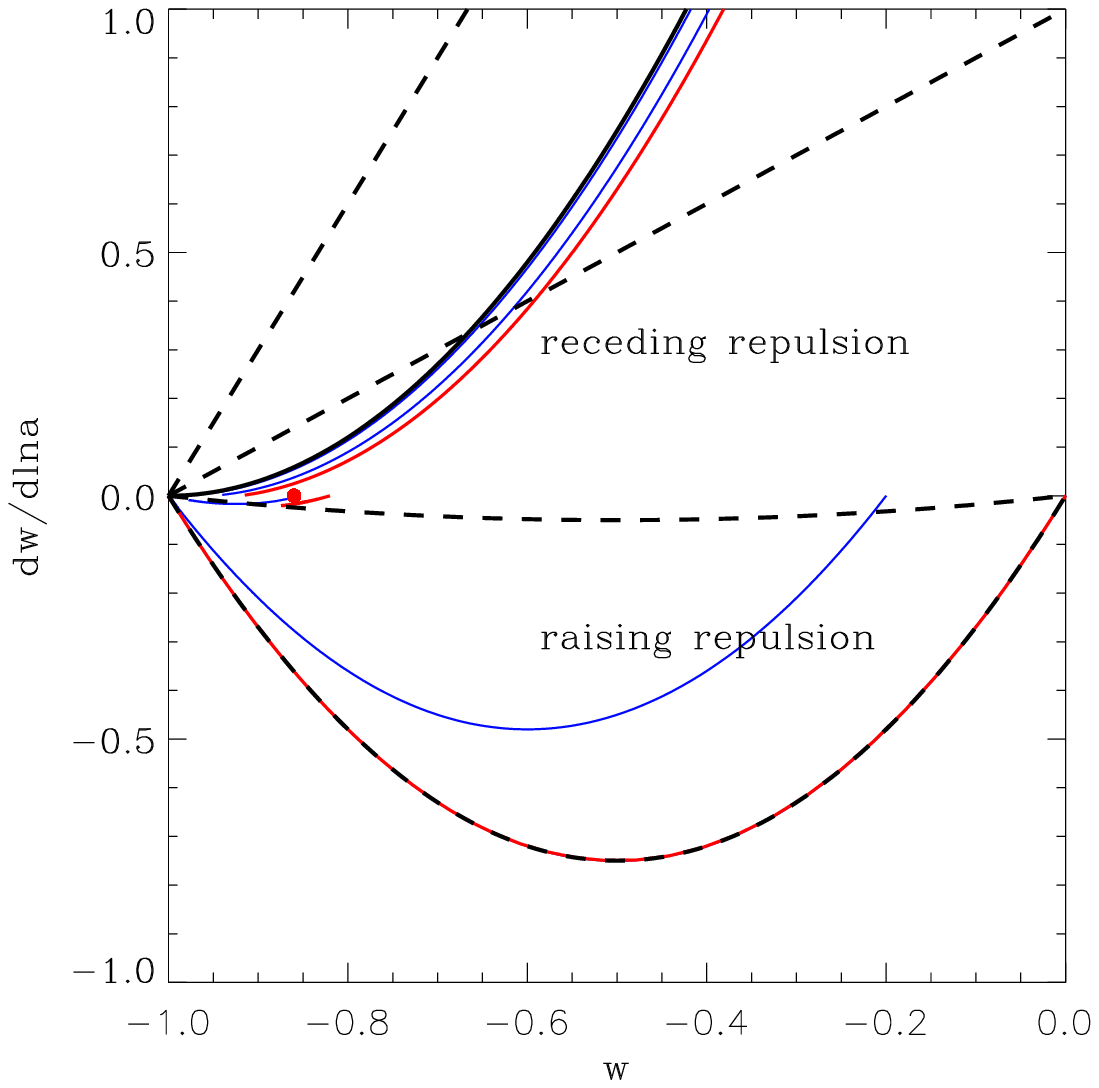}
\caption{The $w-dw/d\ln{a}$ phase plane for scalar fields with barotropic EoS as models of dynamical dark energy (solid lines). If $dw/d\ln{a}<0$ the fields evolve from right to left raising their repulsion properties, if $dw/d\ln{a}>0$ the fields evolve from left to right receding them. Thick dashed lines show the ranges occupied by the thawing and freezing scalar fields deduced by \cite{Caldwell2005} from the analysis of simplest particle physics scalar field models of dynamical dark energy. In the left panel the phase plane evolution tracks of the scalar fields with barotropic EoS are shown in the range $0.5\le a\le 1$ ($0\le z\le 1$) and in the same scale as in \cite{Caldwell2005} for easy comparison. In the right panel the phase plane evolution tracks of the scalar fields with barotropic EoS correspond to the range $0.0001\le a\le 10$ ($-0.9\le z\le 10000$). Thick solid black lines show the limits for such scalar field models: the upper line corresponds to $c_a^2=-1$, the lower one to $c_a^2=0$ (superimposed with the lower limit for freezing scalar fields from \cite{Caldwell2005}). The blue solid lines show the phase tracks of models considered in this section, the red solid lines show the phase tracks of the best fitting models discussed in section \ref{results}. The red point shows the best fitting $w=const$ model.}
\label{w`w}
\end{figure*}

Using the obtained dependences of densities of each component on the scale factor one can deduce from the Einstein equations (\ref{E_eq}) the following equations for background dynamics:
\begin{eqnarray}
 H&=&H_0\sqrt{\Omega_r/a^{4}+\Omega_m/a^{3}+\Omega_k/a^{2}+\Omega_{de}f(a)}, \label{H}\\
 q&=&\frac{1}{2}\frac{2\Omega_r/a^{4}+\Omega_m/a^{3}+(1+3w)\Omega_{de}f(a)}
{\Omega_r/a^{4}+\Omega_m/a^{3}+\Omega_k/a^{2}+\Omega_{de}f_(a)},\label{q}
\end{eqnarray}
where $f(a)=[(1+w_0)a^{-3(1+c_a^2)}+c_a^2-w_0]/(1+c_a^2)$ and $\Omega_k\equiv-K/(aH)^2$ is the dimensionless curvature parameter. $H\equiv\dot{a}/{a^2}$ is the Hubble parameter (expansion
rate) for any moment of time and $q\equiv-\left(a\ddot{a}/\dot{a}^2-1\right)$ is the acceleration
parameter. They completely describe the dynamics of expansion of the homogeneous isotropic Universe.

To analyze the properties of scalar field models of dark energy with barotropic EoS and to calculate the observational predictions the allowed by physics and mathematics ranges of values for $w_0$ and $c_a^2$
should be determined. The field variables $\phi$ (for positive energy density of the scalar field) and $\xi$ are always real  if $w\ge -1$. The potential $U(a)$ of classical scalar field is positive if $w\le 1$ and the potential $\tilde{U}(a)$ of tachyonic one is real if $w\le 0$. As $w(a)$ changes from $c_a^2$ to $w_0$ in the range of $0\le a\le 1$, the allowed ranges of values for $w_0$ and $c_a^2$ are the same for both fields. Other upper constraint for classical scalar field can be obtained from the next speculations. In order to keep the shape of angular power spectrum of CMB temperature fluctuations close to the $\Lambda$CDM one matching well the observational data, the dimensionless density of dark energy $\Omega_{de}(a)=8\pi G\rho_{de}(a)/3H^2(a)$ should not exceed 0.01-0.02 at the last scattering surface. Its asymptotic at small $a$ is $\Omega_{de}(a)\approx \Omega_{de}^{(0)}(1+w_0)a^{-3(1+c_a^2)}/(1+c_a^2)$. The condition $\Omega_{de}(0.001)\le 0.02$ is satisfied for $c_a^2\le 0$. Therefore, for both quantities $w_0$ and $c_a^2$ we accept the ranges of values $-1\le w_0,\;c_a^2\le 0$ for which the scalar field variables and potentials have no singularities during the past history of the Universe, but there is no warranty that the singularities will not appear in future. Let us analyze that.

The system of equations (\ref{w})-(\ref{q}) describes 3 possible variants of future evolution of the Universe defined by the relationship between $w_0$ and $c_a^2$ for given values of rest of the parameters $\Omega_m$, $\Omega_k$, $\Omega_r$ and $\Omega_{de}$ (Fig. \ref{fig1}-\ref{U_X}). 

1) \textbf{$w'<0$ ($c_a^2>w_0$)}: As it follows from (\ref{w}), in this case $w$ decreases monotonically from $c_a^2$ at the early epoch to $w_0$ at current one up to -1 at the infinite time. The constant $C$ in EoS (\ref{beos}) is negative. The dark energy density and pressure tend asymptotically to $\rho_{de}^{(\infty)}=\rho_{de}^{(0)}(c_a^2-w_0)/(1+c_a^2)$ and $P_{de}^{(\infty)}=-\rho_{de}^{(\infty)}$. Therefore, in this case the scalar field rolls slowly to the vacuum (see Fig. \ref{fig1}) and in far future the Universe will proceed into de Sitter stage of its expansion with $q^{(\infty)}=-1$ and $H^{(\infty)}=\sqrt{\Omega_{de}(c_a^2-w_0)/(c_a^2+1)}H_0$.  The scalar fields of such type have the following general properties (see relations (\ref{w_X})-(\ref{pt}) and top panels of Fig. \ref{U_X}): their kinetic terms and potentials have always real positive values; for classical field $X<U$ and for tachyonic one $\tilde{X}<1/2$; their potentials roll slowly to minima ($U_{min},\tilde{U}_{min}>0$), which correspond to the infinite values of the field variables and time; the kinetic terms of these fields tend asymptotically to 0 -- this means that $\dot{\phi}$ and $\dot{\xi}$ $\rightarrow 0$ and the fields will ``freeze''.  

2) {\bf $w'=0$ ($c_a^2=w_0$):} It corresponds to the well-studied case $w=const$. In this case $C=0$ and we have usual barotropic EoS $P_{de}=w_0\rho_{de}$, $\rho_{de}\rightarrow 0$ when $a\rightarrow \infty$. So, the future Universe will experience the power law expansion with $a\propto t^{2/3(1+w_0)}$ and acceleration parameter $q\rightarrow (1+3w_0)/2$. The scalar fields of such type have the following general properties (see relations (\ref{w_X})-(\ref{pt}) and middle panels of Fig. \ref{U_X}): their kinetic terms and potentials have always real positive values; for classical field $U/X=const>1$ and for tachyonic one $\tilde{X}=(1+w_0)/2=const$; their potentials roll very slowly to minima, which correspond to the infinite values of the field variables and time; the potentials of these fields tend asymptotically to 0 -- it means that such fields ``thaw''.  

3) {\bf $w'>0$ ($c_a^2<w_0$):} The EoS parameter $w$ increases monotonically from $c_a^2$ at the early epoch to $w_0$ at the current one and still continues to increase after that. It will reach 0 in future at $a_{(w=0)}=\left[c_a^2(1+w_0)/(c_a^2-w_0)\right]^{\frac{1}{3(1+c_a^2)}}$ when potential of tachyonic scalar field becomes imaginary (relation (\ref{pt}) and right bottom panel of Fig. \ref{U_X}) and 1 at $a_{(w=1)}=\left[\frac{1}{2}(1-c_a^2)(1+w_0)/(w_0-c_a^2)\right]^{\frac{1}{3(1+c_a^2)}}$ 
when potential of the classical one becomes negative (relation (\ref{pcl}) and left bottom panel of Fig. \ref{U_X}). The scalar field densities at these $a$ are positive: $\rho_{de}(a_{(w=0)})=\rho_{de}^{(0)}(c_a^2-w_0)/c_a^2$ and $\rho_{de}(a_{(w=1)})=\rho_{de}^{(0)}(c_a^2-w_0)/(c_a^2-1)$ correspondingly. The field will satisfy the strong energy condition $\rho_{de}+3P_{de}\ge 0$ starting from
$a= \left[(1+w_0)(1+3c_a^2)/(2(c_a^2-w_0))\right]^{\frac{1}{3(1+c_a^2)}}$ and then accelerated expansion of the Universe will be changed by the
decelerated one. The density of scalar field continues decreasing, reaches 0 at 
$a_{(\rho=0)}=\left[(1+w_0)/(w_0-c_a^2)\right]^{\frac{1}{3(1+c_a^2)}}$ and then
becomes negative. The EoS parameter at this moment will have discontinuity of the second kind (Fig. \ref{fig1}). Later, when $\rho_m+\rho_{de}$ reaches 0, the expansion of the Universe will be changed by the contraction since at this moment $\dot{a}=0$, $\ddot{a}<0$, as it follows from equations (\ref{H}) and (\ref{q}), which have no solution for larger $a$. Such behavior can be corrected only slightly by the curvature parameter from the allowable range. Though the potential and field variable of tachyonic scalar field become imaginary at $a_{(w=0)}<a<a_{(\rho=0)}$ and $a>a_{(\rho=0)}$ correspondingly (their energy density and pressure are real value always) we have no substantial arguments for ruling out such models since they have not singularities in past and are not excluded by the observational data. We have also no physical argument against the possibility of generation of such fields at the early epoch. Meanwhile, we will use the condition $c_a^2>w_0$ too for analysis of the scalar field dark energy without singularities in future. The appearance of imaginary values of either field variable or potential of the tachyonic scalar field suggests that this case needs to be analyzed using the multifield model of dark energy. That will be the matter of a separate paper.
\begin{figure*}
\includegraphics[height=.47\textheight]{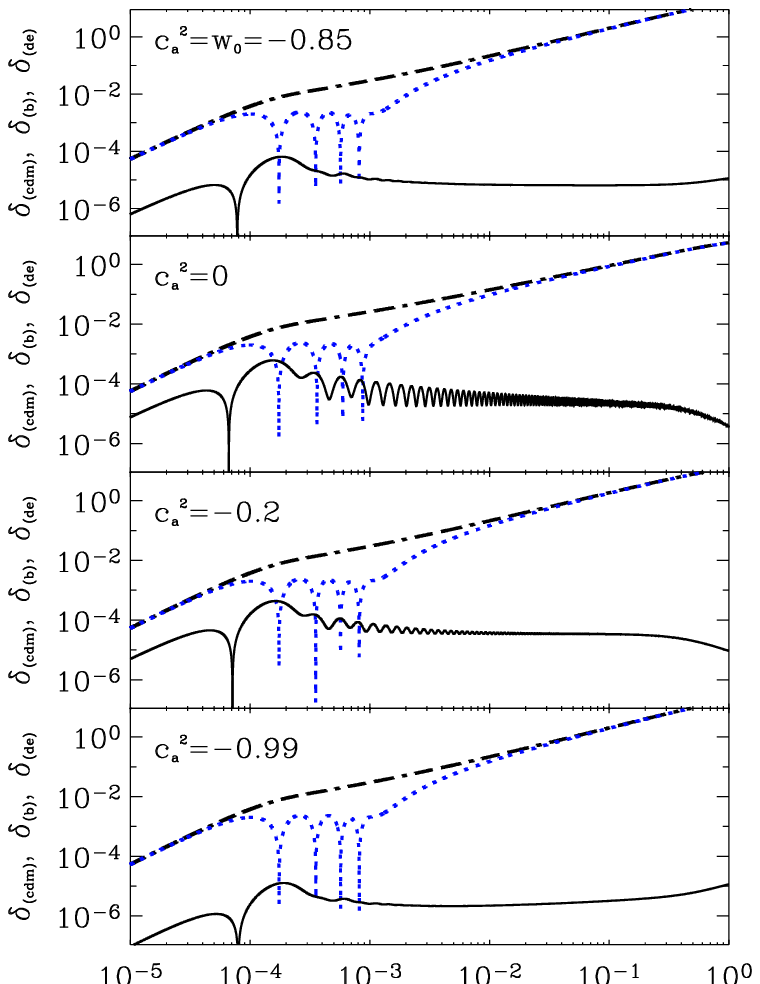}
\includegraphics[height=.47\textheight]{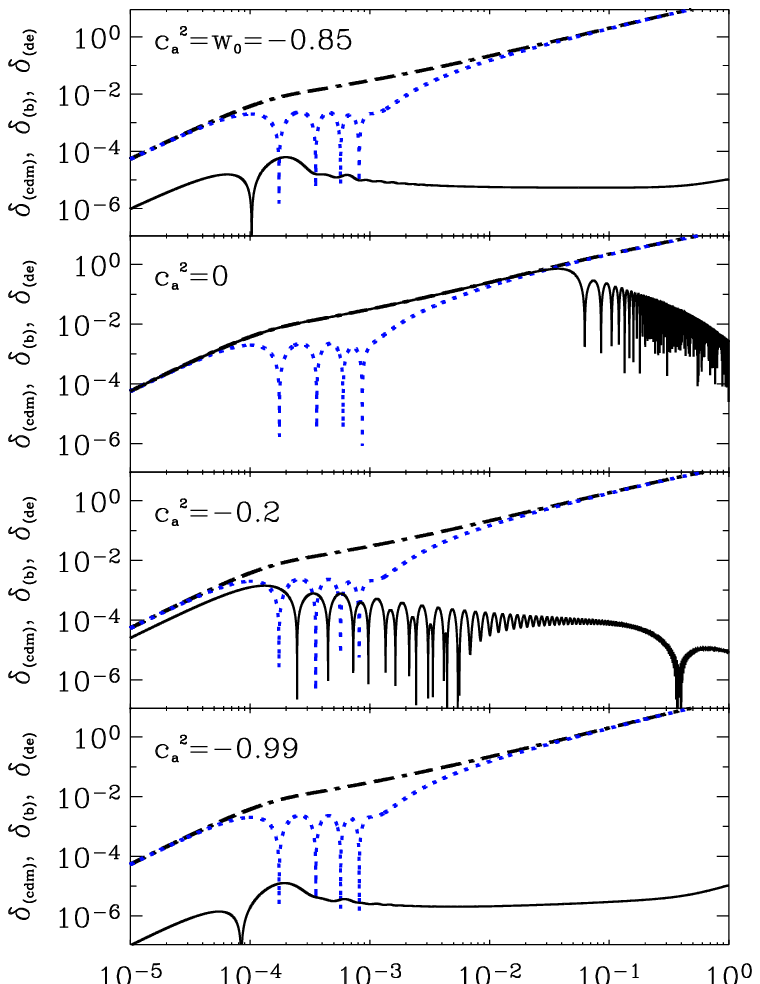}
\caption{Evolution of linear density perturbations of cold dark matter (dashed line), baryons (dotted) and scalar field (solid) with $c_a^2=w_0=-0.85$, $c_a^2=0$, $c_a^2=-0.2$, $c_a^2=-0.99$ (from top to bottom). Classical scalar field - left column, tachyonic one - right. The wave number of perturbations is $k=0.1$ Mpc$^{-1}$. }
\label{ddeb}
\end{figure*}

At the end of the analysis of general properties of the scalar field models of dark energy with $c_a^2=const$ let us examine their occupancy of the $w-dw/d\ln a$ phase plane. From (\ref{w'}) and the constraint $-1\le c_a^2\le 0$ follows that the scalar field models of dark energy with $c_a^2=const$ occupy the $w-dw/d\ln a$ region limited by the lines $dw/d\ln a=3(1+w)^2$ and $dw/d\ln a=3w(1+w)$ (Fig. \ref{w`w}). The last one coincides with the lower limit for freezing scalar field models of dark energy deduced by \cite{Caldwell2005} from the analysis of the simplest particle-physics models of cosmological scalar fields. Below it the scalar fields have too large density at the early epoch that contradicts the data on CMB anisotropy. Above the upper limit there is a range of fields which started as phantom ones, which is excluded for fields with classical and tachyonic Lagrangians considered here. The scalar fields which are in the phase plane between the lines $dw/d\ln a=0$ and $dw/d\ln a=3w(1+w)$ evolve from right to left in Fig. \ref{w`w} and their repulsion properties are raising with time. They are unlimited in time and $w$ for them tends asymptotically to -1. The scalar fields which are in the phase plane between the lines $dw/d\ln a=0$ and $dw/d\ln a=3(1+w)^2$ evolve from left to right in the Fig. \ref{w`w} and their repulsion properties are receding with time ($dw/d\ln a>0$, $w$ increases) to change the accelerated expansion by decelerated one and even collapse. They can start in the range below the lower limit for thawing scalar fields, then enter the range of thawing scalar fields limited by \cite{Caldwell2005}, cross it and go out of upper limit $dw/d\ln a=3(1+w)$ when $w>0$. So, the  scalar fields with $dw/d\ln a>0$ ($c_a^2<w_0$) can  only partially be called thawing. We propose to call them ``scalar fields receding repulsion'', reflecting their main properties. Symmetrically, the  scalar fields with $dw/d\ln a<0$ ($c_a^2>w_0$), occupying the same range as freezing scalar fields from \cite{Caldwell2005}, can be called ``scalar fields raising repulsion''. Establishing of how well they can fit the current observational data is the goal of the paper.

Since the classical scalar fields are completely indistinguishable from the tachyonic ones when considering only the background dynamics, the linear theory of perturbations in the multicomponent Universe must be included into analysis.

\section{Evolution of linear perturbations in the synchronous gauge}\label{evolveeq}
\begin{figure*}
\includegraphics[width=.47\textwidth]{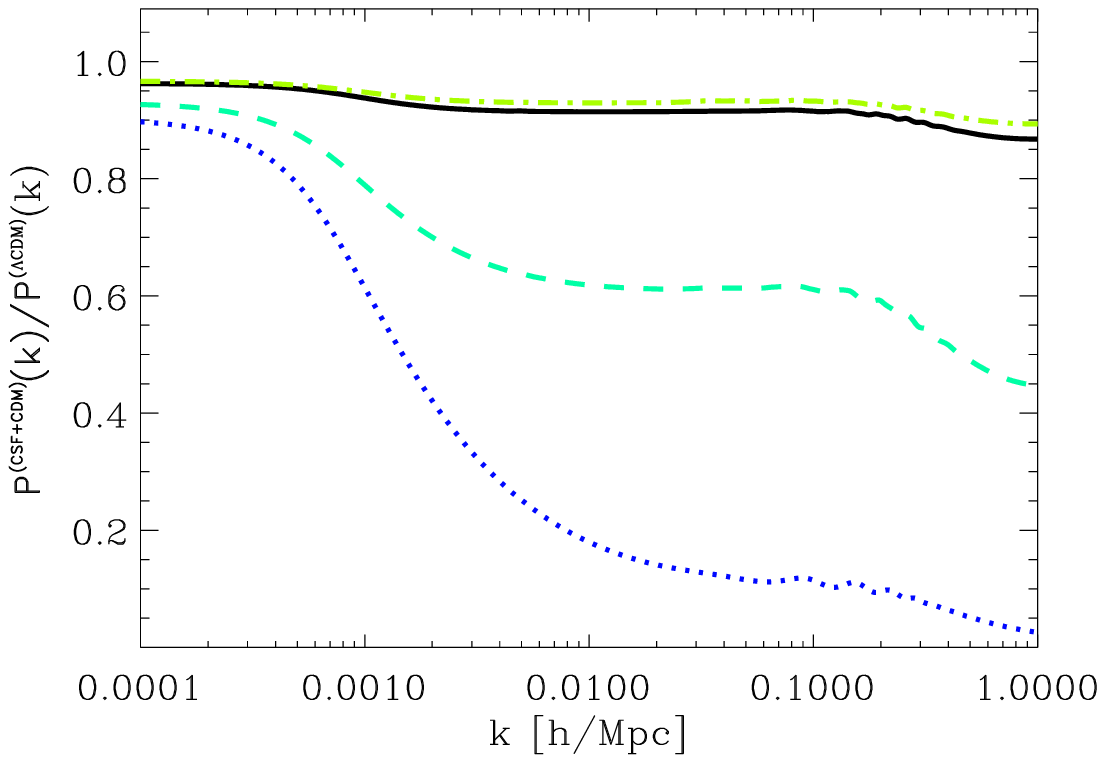}
\includegraphics[width=.47\textwidth]{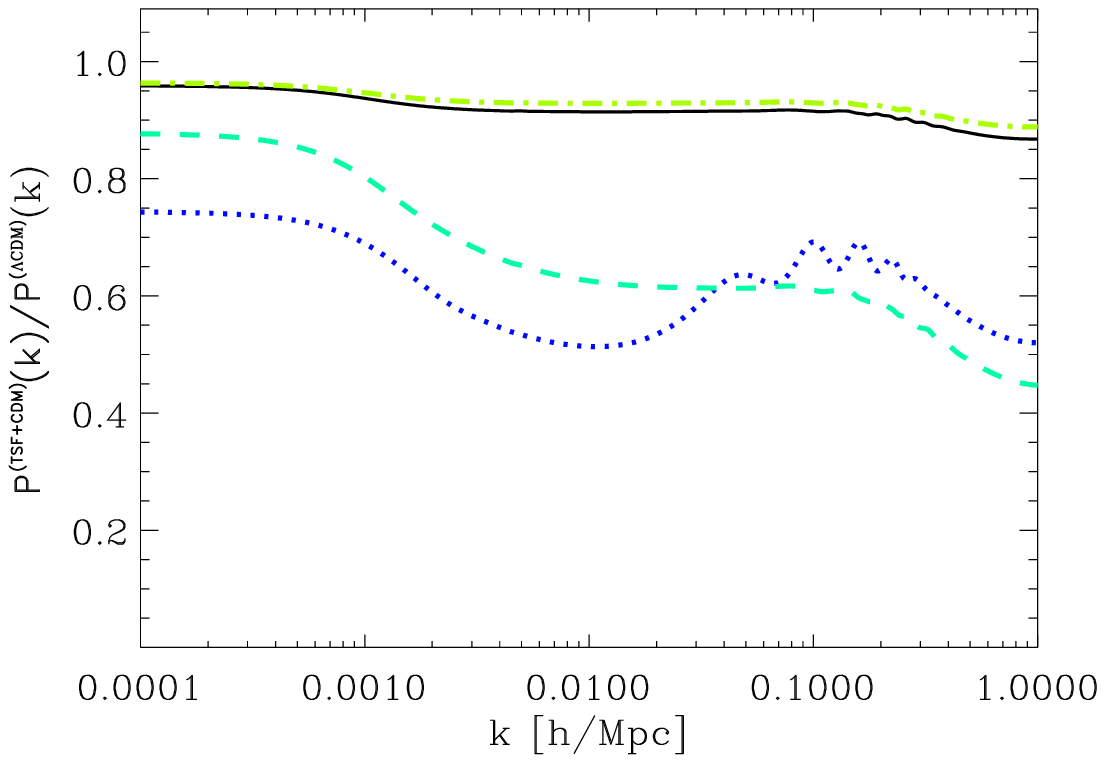}
\includegraphics[width=.47\textwidth]{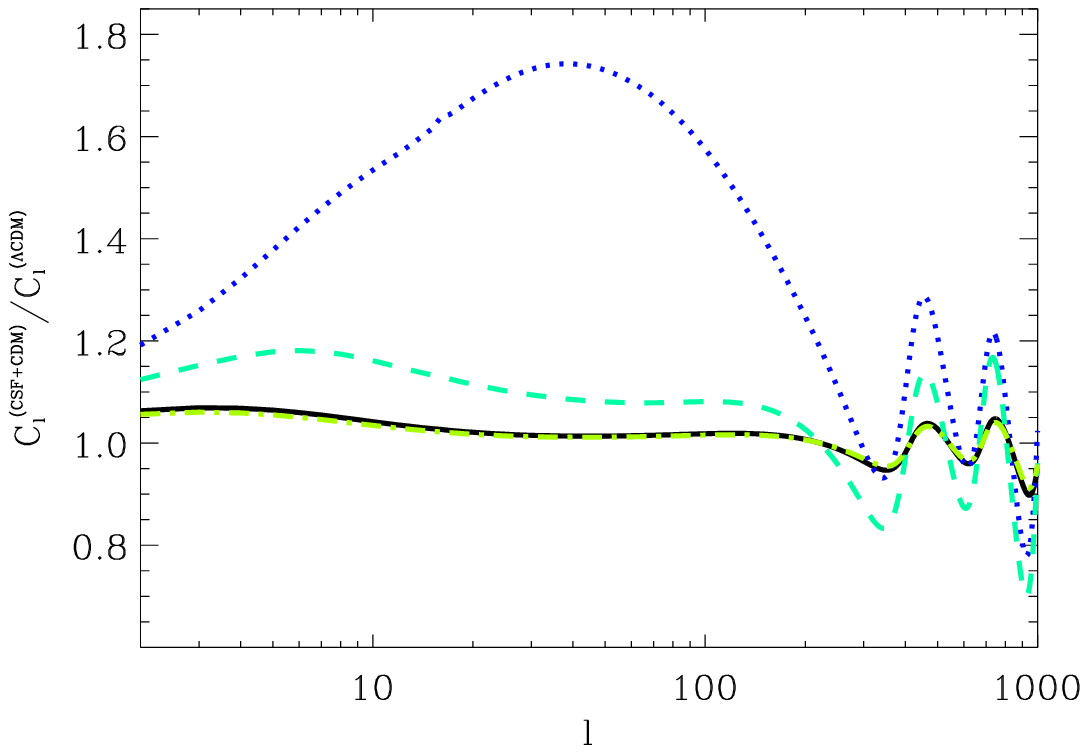}
\includegraphics[width=.47\textwidth]{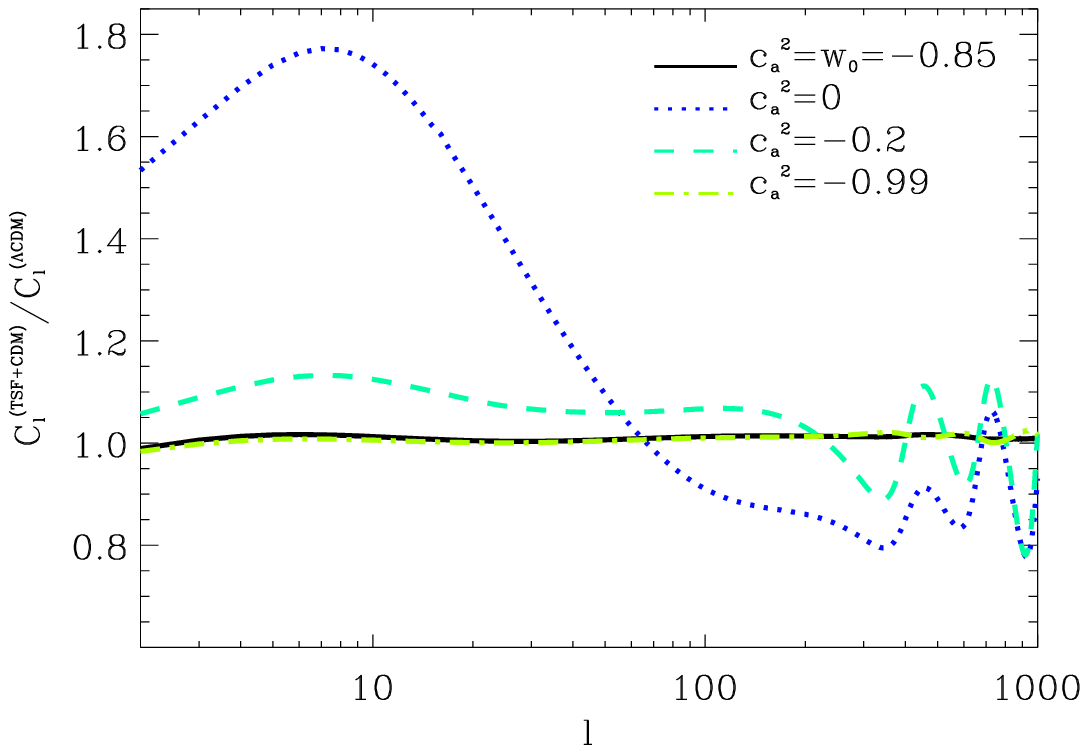}
\caption{Top panels: The ratios of matter density power spectra $P(k)$ in the models with classical (left) and tachyonic (right) scalar fields to the matter density power spectra in the $\Lambda$CDM model. Bottom panels: the ratios of CMB angular power spectra $C_l$ in the models with classical (left) and tachyonic (right) scalar fields to the corresponding ones in the $\Lambda$CDM model. In each panel the solid line stands for the scalar field with $c_a^2=w_0=-0.85$, dotted one for  $c_a^2=0$, dashed one for $c_a^2=-0.2$, dash-dotted for $c_a^2=-0.99$.}
\label{de2l}
\end{figure*}

Now we consider the spatially flat ($\Omega_k=0$) background with scalar perturbations in the synchronous gauge with the line element
\begin{eqnarray}
ds^2=g_{ij} dx^i dx^j =a^2(\eta)(d\eta^2-(\delta_{\alpha\beta}+h_{\alpha\beta}) dx^{\alpha}dx^{\beta}).\label{ds}
\end{eqnarray}
The scalar perturbations of metric $h_{\alpha\beta}$ can be decomposed into the trace $h\equiv h_{\alpha}^{\alpha}$ and traceless $\tilde{h}_{\alpha\beta}$ components as $h_{\alpha\beta}=h\delta_{\alpha\beta}/3+\tilde{h}_{\alpha\beta}$. The perturbations are supposed to be small ($h\ll1$), henceforth all following equations are linearized with respect to the perturbed
variables. In the multicomponent fluid each component moves with a small peculiar velocity $V^{\alpha}=dx^{\alpha}/d\eta$, defined by its intrinsic properties (density, pressure, entropy etc.) and $h$. At the linear stage of evolution of perturbations the cold dark matter (CDM) component is a pressureless perfect fluid interacting with other components only via gravity. Therefore, the synchronous coordinates are usually defined as comoving to the particles of CDM.
\begin{figure*}
\includegraphics[width=.48\textwidth]{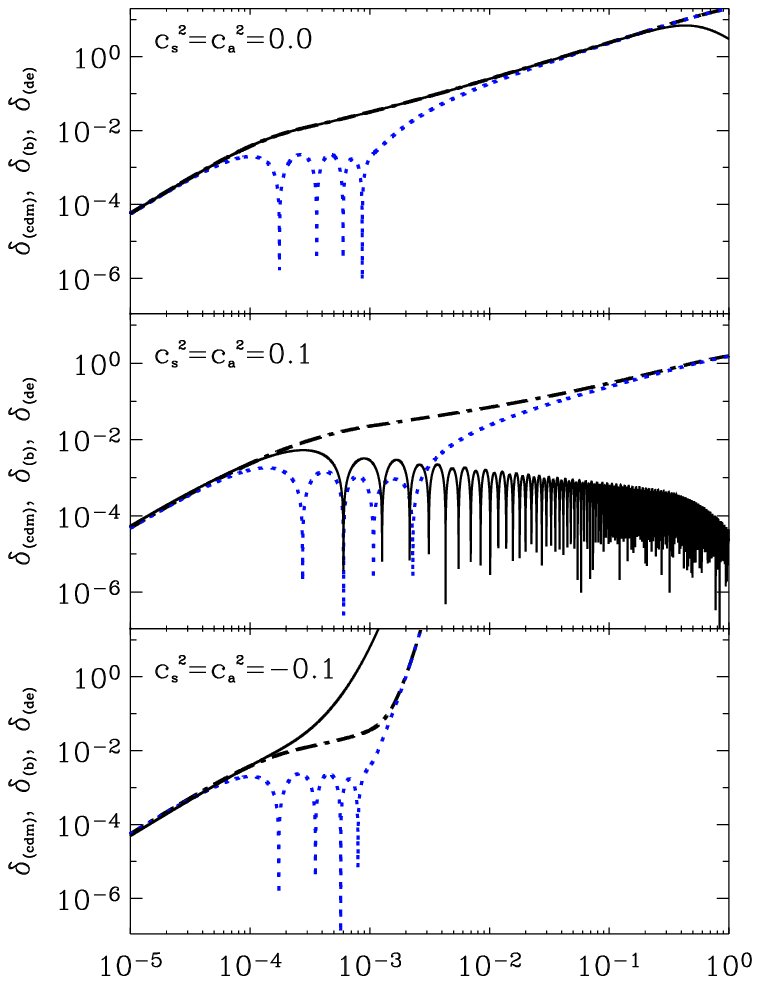}
\includegraphics[width=.48\textwidth]{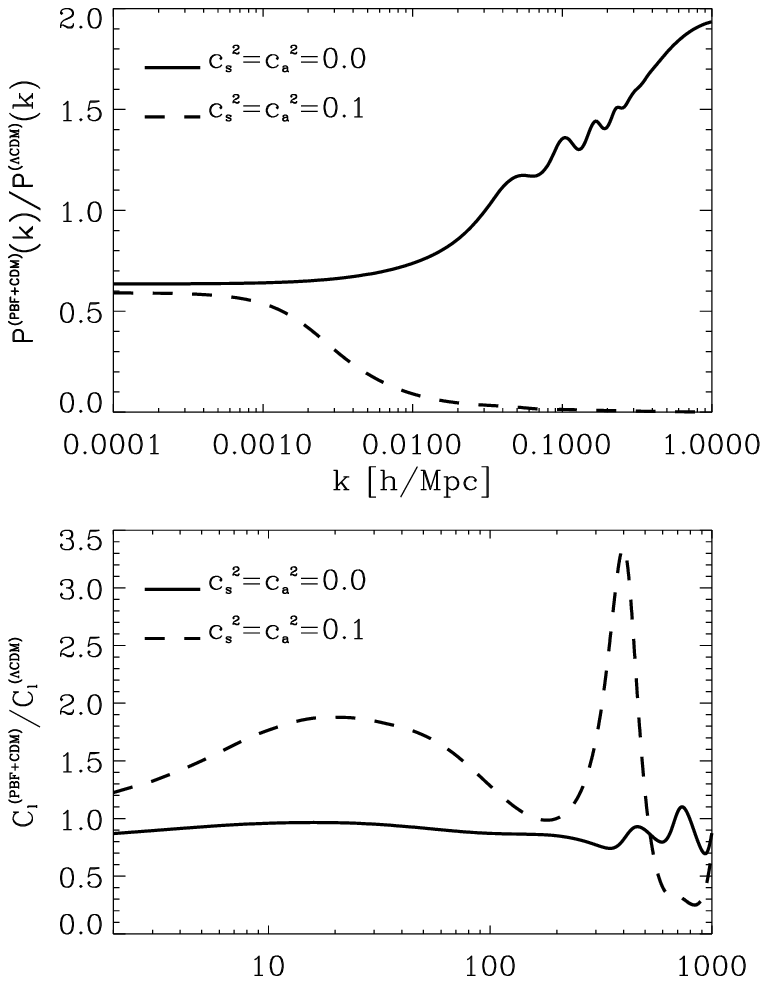}
\caption{Pure barotropic fluid models of dark energy in the multicomponent Universe. Left: the evolution of linear density perturbations ($k=0.1$ Mpc$^{-1}$) of cold dark matter (dashed line), baryons (dotted) and barotropic fluid (solid) with $c_s^2=c_a^2=0.0$, $c_s^2=c_a^2=0.1$  and $c_s^2=c_a^2=-0.1$ (from top to bottom). Right: the ratios of matter density power spectrum $P(k)$ (top panel) and CMB angular power spectrum $C_l$ in the model with barotropic fluid to the corresponding ones in the $\Lambda$CDM model. In each panel the solid line corresponds to the pure barotropic fluid with $c_s^2=c_a^2=0.0$, dashed one -- with $c_s^2=c_a^2=-0.1$.}
\label{ddeb_pbl}
\end{figure*}
The energy density $\rho_{(n)}$ and pressure $P_{(n)}$ of each component are perturbed as $\rho_{(n)}(x^{\alpha},\eta)=\overline\rho_{(n)}(\eta)(1+\delta_{(n)}(x^{\alpha},\eta))$ and $P_{(n)}(x^{\alpha},\eta)=\overline P_{(n)}(\eta)+\delta P_{(n)}(x^{\alpha},\eta)$. They are measured by a comoving observer being at rest relative to the fluid at the instant of measurements, so their perturbations are related as $\delta P_{(n)}=c_{s\,(n)}^2\overline\rho_{(n)}\delta_{(n)}$, where $c_{s\,(n)}^2$ is the comoving effective sound speed of component $(n)$. This sound speed equals to zero for dust medium (e. g. CDM or baryonic matter after recombination at scales above Jeans scale) and $1/3$ for relativistic components. For scalar fields the entropy perturbations are inherent and cause in addition to the adiabatic pressure perturbations, which follow from the variation of (\ref{beos}), the non-adiabatic pressure ones $\delta P_{(de)}^{(nad)}$, so the total perturbation is their sum \cite{deltap}: 
\begin{eqnarray}
\delta P_{(de)}=c_a^2\delta \rho_{(de)}+\delta P_{(de)}^{(nad)}.\nonumber
\end{eqnarray}
In the rest frame of dark energy ($V_{(de)}=\delta\phi=\delta\xi=0$) it can be presented as 
\begin{eqnarray}
\delta P_{(de)}=c_{s\,(de)}^2\rho_{(de)}\delta_{(de)},\nonumber
\end{eqnarray}
where the effective (rest-frame) sound speed $c_{s\,(de)}^2$ for the scalar field with given Lagrangian can be calculated as
\begin{eqnarray}
c_{s\,(de)}^2\equiv\frac{P_{,X}}{\rho_{,X}}=\frac{L_{,X}}{2XL_{,XX}+L_{,X}}.\nonumber
\end{eqnarray}
It equals 1 for the classical scalar field and $-w$ for the tachyonic one. 

For pure barotropic fluid\footnote{The conditions on the Lagrangian for vanishing non-adiabatic pressure perturbation in the case of generic scalar field can be found in \cite{Unnikrishnan2010}.} $\delta P_{(de)}^{(nad)}=0$ and $c_s^2=c_a^2$ -- that distincts it from the scalar fields with barotropic equation of state.

In the linear perturbation theory it is convenient to perform the Fourier transformation of all spatially-dependent variables and use the equations for corresponding Fourier amplitudes. The differential energy-momentum conservation law $\delta T^{i\;\;(de)}_{j\;;i}=0$ for perturbations in the space with metric (\ref{ds}) gives the equations for evolution of Fourier amplitudes of density and velocity perturbations of dark energy in the synchronous gauge:
\begin{eqnarray}
&&\dot{\delta}_{(de)}+3(c_s^2-w)aH\delta_{(de)}+(1+w)\frac{\dot{h}}{2}\nonumber\\ 
&&+(1+w)\left[k+9a^2H^2\frac{c_s^2-c_a^2}{k}\right]V_{(de)}=0, \label{d_de}\\
&&\dot{V}_{(de)}+aH(1-3c_s^2)V_{(de)}-\frac{c_s^2k}{1+w}\delta_{(de)}=0.\label{V_de}
\end{eqnarray}
These equations in the conformal-Newtonian gauge are presented in \cite{Putter2010}, in the gauge-invariant variables in \cite{Sergijenko2009b}. In these papers it was shown also that the perturbations of scalar fields with different $c_s^2$ evolve differently. For the case of classical and tachyonic scalar fields with $c_a^2=0$ the evolution was extensively studied for the two-component (dark energy plus dark matter) model in \cite{Sergijenko2009b}. It was shown also that the transfer function of the cold dark matter perturbations is different for the same set of cosmological parameters but different Lagrangians (and thus $c_s^2$) of dark energy.

The equations for the rest of components (non-relativistic and relativistic) are the same as in \cite{Ma1995}, so we skip this part here. For calculation of the evolution of perturbations in all components and power spectra of matter density perturbations and cosmic microwave
anisotropy we have used the publicly available code CAMB
\cite{camb,camb_source}, modified to include the presented here expressions for $H(a)$, $\rho_{(de)}(a)$ and the evolution equations for dark energy perturbations. The evolution of density perturbations ($k$=0.1 Mpc$^{-1}$) of cold dark matter, baryons and scalar fields with different $c_a^2$ (0, -0.2, -0.85, -0.99) is shown in Fig. \ref{ddeb}. Other cosmological parameters have the next values: $\Omega_{de}=0.722$, $w_0=-0.85$, $\Omega_{cdm}h^2=0.111$, $\Omega_bh^2=0.0227$, $H_0=66.1$ km/s/Mpc, $n_s=0.975$, $\tau_{rei}=0.085$. The adiabatic initial conditions from \cite{Ma1995} for all components except dark energy are used here and below. The initial conditions for dark energy perturbations are obtained from the asymptotic solutions of (\ref{d_de})-(\ref{V_de}) for $k\eta\ll1$ (the superhorizon perturbations) in the early radiation-dominated epoch and are following:
\begin{eqnarray}
&&\delta_{(de)}^{\;init}=-\frac{(4-3c_s^2)(1+w)}{8+6c_s^2-12w+9c_s^2(w-c_a^2)}h,\\
&&V_{(de)}^{\;init}=-\frac{c_s^2k\eta_{init}}{8+6c_s^2-12w+9c_s^2(w-c_a^2)}h.\label{v_de_init}
\end{eqnarray}
The character of evolution of scalar field density perturbations depends on the temporal behavior of EoS parameter, the difference of adiabatic and effective sound speeds (the value of intrinsic entropy of scalar field is proportional to it) and the ratio of the scale of perturbation to particle horizon (see equations (\ref{d_de})-(\ref{V_de})). We can see that the magnitude of density perturbations of such fields increases before entering the  
acoustic horizon, decreases and changes the sign during entering one and decays with oscillations later. So, at the current epoch the magnitude of dark 
energy linear density perturbations is essentially lower than matter one at all subhorizon scales. 

Evolution of matter density perturbations is affected by the dynamics of background expansion through the scale-independent expansion rate and by scalar field density perturbations through the perturbations of metric which depend on scale. For visualization of such total influence of dark energy on matter density evolution we present in Fig. \ref{de2l} the ratios of matter density power spectrum $P(k)$ and CMB angular power spectrum $C_l$ in the models with scalar field to the corresponding ones in the $\Lambda$CDM model with the same cosmological parameters. The dependence of both ratios on scale, $c_a^2$ and type of scalar field inspires hope that usage of the accurate enough dataset will give the possibility to constrain the scalar field models of dark energy considered here. For that we use here the available currently observational data. 

At the end of this section let us discuss the difference between scalar field models of dark energy with barotropic equation of state and pure barotropic fluid models of dark energy. For pure barotropic fluid (PBF) $c_s^2=c_a^2$. The analysis of equation (49) from \cite{Sergijenko2009b} shows that such dark energy is strongly gravitationally instable at subhorizon scales of density perturbations ($k\eta\ll 1$) for $c_s^2<0$. So, such fluid model of dark energy will not contradict the observational data when $c_s^2=c_a^2>0$. In the left panel of Fig. \ref{ddeb_pbl} the evolution of linear density perturbations of cold dark matter, baryons and pure barotropic fluid with $c_s^2=c_a^2=0$, $c_s^2=c_a^2=0.1$  and $c_s^2=c_a^2=-0.1$ is shown for perturbations with the wave number $k=0.1$ Mpc$^{-1}$. The curves for last case ($c_s^2=-0.1$) illustrate the strong instabilities in perturbation growth, so that the models with $c_s^2<0$ should be avoided. In order to check the viability of PBF models of dark energy we have computed the ratios of matter density power spectrum $P(k)$ and CMB angular power spectrum $C_l$ in such models to the corresponding ones in the $\Lambda$CDM model. In the right panel of Fig. \ref{ddeb_pbl} one can see, that the influence of PBF dark energy with $c_s^2=c_a^2=0$ on the evolution of matter density perturbations differs from one of classical and tachyonic scalar fields with $c_a^2=0$ (see Fig. \ref{de2l}). The large difference between the power spectra of PBF+CDM with $c_s^2=0.1$ and $\Lambda$CDM models means that the upper observational constraint on $c_s^2$ should be close to zero. 

In the context of this discussion and results, presented in this section, we would like to explain the title of the analyzed here dark energy model -- scalar field with barotropic equation of state. It was shown above that the scalar field is not exactly barotropic because of the non-adiabaticity, which appears at the level of linear perturbations and is $\propto \delta_{(de)}$. Since $\delta_{(de)}\ll 1$ now and in the past at all scales (at large scales due to the initial power spectrum $\propto k^{n_s}$, at small scales due to the decaying of subhorizon perturbations of the scalar field, shown in Fig. \ref{ddeb}), in the main order the dark energy equation of state is the generalized linear barotropic one. This property defines the scalar field dynamics as well as the dynamics of expansion of the Universe and its future. That is why we call such scalar field the ``scalar field with barotropic equation of state''.

\section{Method and data}\label{method_data}
\begin{figure*}
\centerline{\includegraphics[width=.32\textwidth]{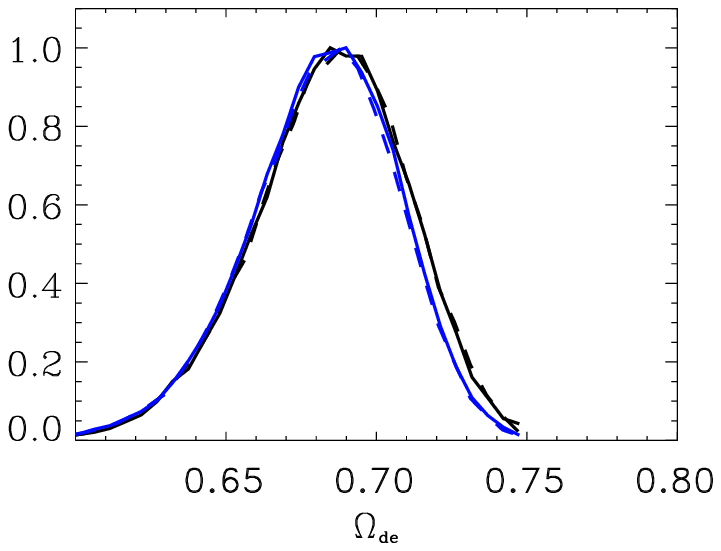}
\includegraphics[width=.32\textwidth]{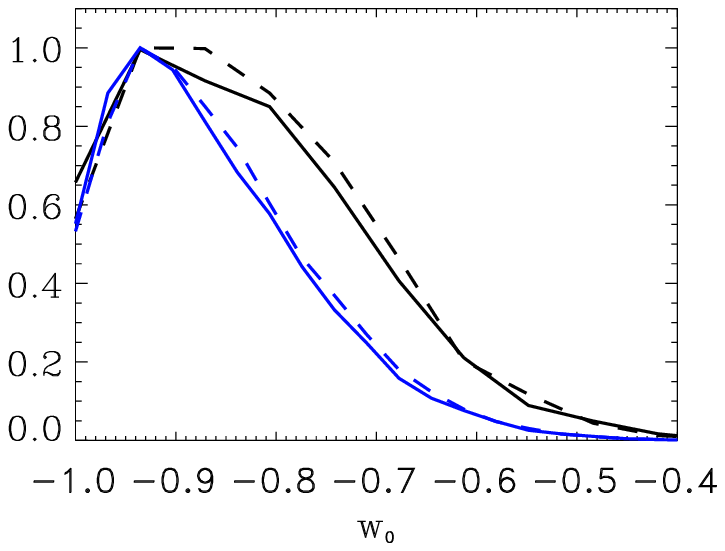}
\includegraphics[width=.32\textwidth]{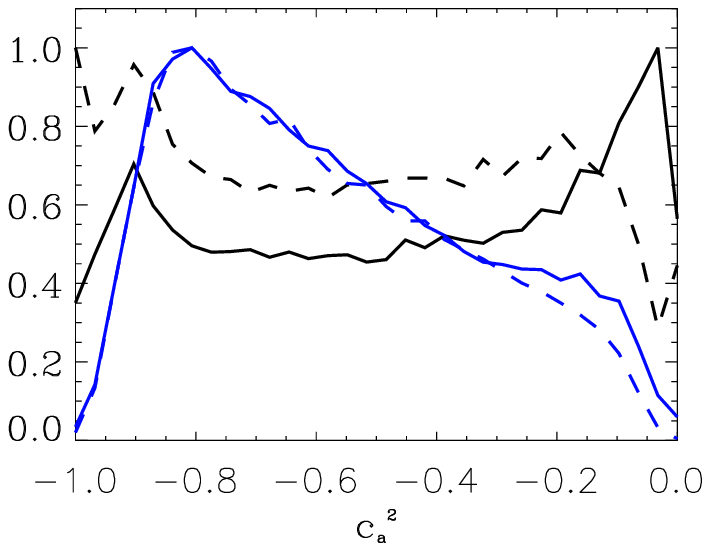}}
\centerline{\includegraphics[width=.32\textwidth]{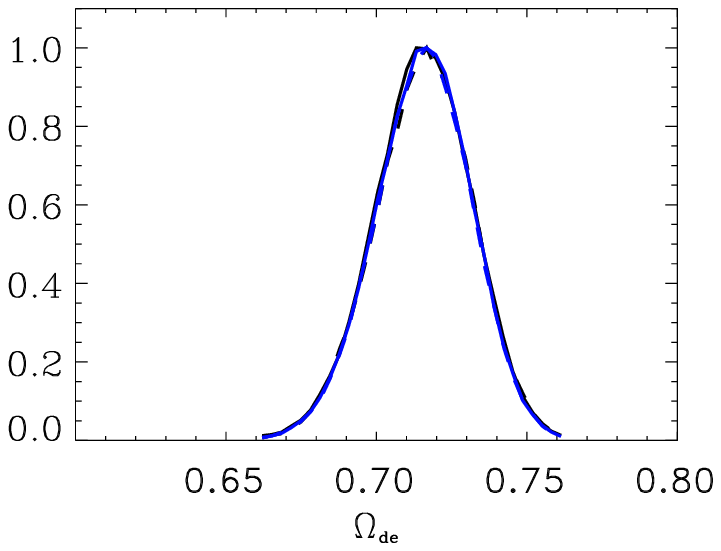}
\includegraphics[width=.32\textwidth]{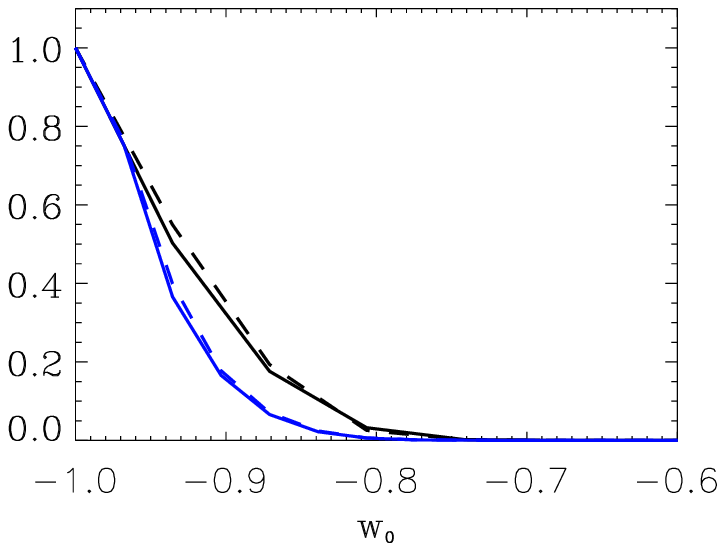}
\includegraphics[width=.32\textwidth]{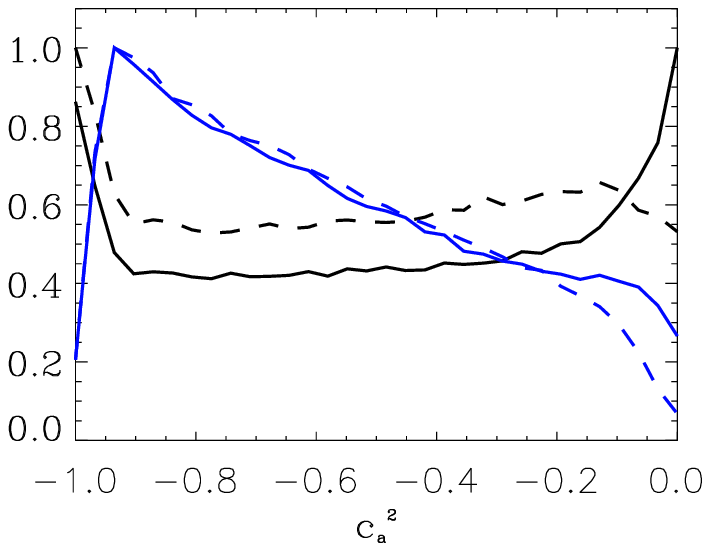}}
\caption{One-dimensional marginalized posteriors (blue lines) and mean likelihoods (black ones, averaged in each of 30 bins) for $\Omega_{de}$, $w_0$ and $c_a^2$ (from left to right) for the combined datasets WMAP7+SDSS LRG DR7 (upper row) and WMAP7+SDSS LRG DR7+SN+HST+BBN (bottom row). Solid lines correspond to the classical scalar field, dashed lines -- to the tachyonic one.}
\label{likestats}
\end{figure*}
To determine the best fitting values and confidential ranges of the  scalar field parameters  together with other cosmological ones we perform the Markov Chain Monte Carlo (MCMC) analysis for the set of current observational data, which include the power spectra from WMAP7 \cite{wmap7} and SDSS DR7 \cite{Reid2009}, the Hubble constant measurements \cite{Riess2009}, the light curves of SNIa \cite{SNUnion2} and Big Bang nucleosynthesis (BBN) prior \cite{bbn}.  
 
We use the publicly available package CosmoMC \cite{cosmomc,cosmomc_source}, which includes the code CAMB  \cite{camb,camb_source} for calculation of model predictions for sampled sets of 8 cosmological parameters\footnote{The curvature parameter was fixed at $\Omega_K=0$, so $\Omega_{de}=1-\Omega_b-\Omega_{cdm}$.} (for recent usage of this package see \cite{cosmomc_recent}):  the present value of the dark energy EoS $w_0$, the physical energy density of cold dark matter $\Omega_c h^2$, the physical energy density of baryons $\Omega_b h^2$, the Hubble constant\footnote{Instead of $H_0$ the CosmoMC code varies the parameter $\theta$ which is the ratio of the sound horizon to the angular diameter distance multiplied by 100.} $H_0$, the scalar spectral index of the primordial perturbations $n_s$, the amplitude of primordial perturbations $A_s$, the reionization optical depth $\tau$ and the Sunyaev-Zel'dovich amplitude $A_{SZ}$. The run of original CosmoMC with dataset listed above gives the best fitting values and confidential ranges for these parameters which are presented in the first and second columns of Table \ref{tabl2}.

The original CAMB has been modified as described above, so the CosmoMC has been run with proposed here parametrization of dark energy EoS parameter (\ref{w}). The extra parameter $c_a^2$ can be either fixed or left free for determination of its best fitting value. Thus, we have 9 free parameters to be explored
with MCMC. By now, we have excluded the phantom models ($w_{de}<-1$) and assumed flat priors for $w_0$ and $c_a^2$ in the range $-1<w_0,\,c_a^2\le 0$.

\section{Results and discussion}\label{results}

\begin{table}
 \caption{The best fitting values and 1$\sigma$
 confidential ranges of cosmological parameters in the
 CSF+CDM and TSF+CDM models, determined on the base
 of WMAP7 and SDSS LRG DR7 data. The current
  Hubble parameter $H_0$ is in units km$\,$s$^{-1}\,
 $Mpc$^{-1}$, the age of the Universe $t_0$ is given 
 in Giga years.}
 \medskip
 \begin{tabular}{c|c|c}
 \hline
 & & \\
 Parameters&CSF+CDM&TSF+CDM \\
 \hline
 & & \\
$\Omega_{de}$&0.69$_{-0.08}^{+0.06}$&0.68$_{-0.08}^{+0.06}$\\
 & & \\
$w_0$&-0.80$_{-0.20}^{+0.34}$&-0.95$_{-0.05}^{+0.46}$\\
 & & \\
$c_a^2$&-0.91$_{-0.09}^{+0.91}$&-0.20$_{-0.80}^{+0.19}$\\
 & & \\
$100\Omega_b h^2$&2.27$_{-0.15}^{+0.16}$&2.27$_{-0.16}^{+0.17}$\\
 & & \\
$10\Omega_{cdm} h^2$&1.10$_{-0.13}^{+0.14}$&1.12$_{-0.15}^{+0.12}$\\
 & & \\
$H_0$&65.4$_{- 7.1}^{+ 7.0}$&65.0$_{- 7.1}^{+ 7.4}$\\
 & & \\
$n_s$&0.97$_{-0.04}^{+0.05}$&0.97$_{-0.04}^{+0.04}$\\
 & & \\
$\log(10^{10}A_s)$&3.08$_{-0.10}^{+0.09}$&3.08$_{-0.10}^{+0.09}$\\
 & & \\
$z_{rei}$&10.8$_{- 3.6}^{+ 3.2}$&10.4$_{- 3.1}^{+ 3.3}$\\
 & & \\
$\sigma_{8}$&0.75$_{-0.21}^{+0.13}$&0.74$_{-0.21}^{+0.15}$\\
 & & \\
$t_0$&13.8$_{- 0.3}^{+ 0.6}$&14.0$_{- 0.5}^{+ 0.5}$\\
 & & \\
 \hline
$-\log L$&3760.26&3760.39\\
 \hline
 \end{tabular}
 \label{tabl1}
 \end{table}

We have performed MCMC runs for 9 spatially flat cosmological models with the dataset described above. All models consist of 5 components: scalar field as dark energy, dark matter, baryons, thermal background radiation and massless active neutrinos. The models differ by the type of scalar field (classical -- CSF or tachyonic -- TSF) and additional conditions for values of adiabatic sound speed ($c_a^2=w_0$, $c_a^2=0$, $c_a^2>w_0$ and $c_a^2$ free). We assume that dark matter is cold (CDM) and since the density of dark components are dominating now and defining the dynamical and clustering properties of our Universe the models are called CSF+CDM or TSF+CDM with corresponding additional conditions for $c_a^2$ which determine the type of dark energy. Each run has 8 chains and the number of samples in each chain is $\sim200\,000$. The main results are presented in the Tables \ref{tabl1} and \ref{tabl2}, where the parameter values of best fit samples and 1$\sigma$ confidential ranges are shown for the main cosmological parameters. In the last row of the tables the quantity $-\log{L}=\chi^2/2$ of best fit sample is presented for each model.

The most interesting models with free $c_a^2$ we have tested using 2 sets of observational data: WMAP7+SDSS LRG DR7 (Table \ref{tabl1}) and WMAP7+SDSS LRG DR7+SN+HST+BBN (Table \ref{tabl2}, last 2 columns). The normalized mean likelihood functions $L(x_j;\theta_i)$ and marginalized posterior distributions $\tilde{L}(\theta_i;x_j)$ for $\Omega_{de}$, $w_0$ and $c_a^2$ are presented for both datasets in Fig. \ref{likestats}. The first of them, $L(x_j;\theta_i)$, is the probability distribution of the observational data $x_j$ for given model parameters $\theta_i$. The second one, $\tilde{L}(x_j;\theta_i)$, is the probability distribution of the parameters $\theta_i$ for given observed data $x_j$.

We can see that for both fields the current density $\Omega_{de}$ is constrained well by both dataset: the likelihoods and posteriors are close. The current density $\Omega_{de}$ and EoS parameters $w_0$ are determined more accurately by the complete dataset: $\Omega_{de}=0.72_{-0.06}^{+0.04}$, $w_0=-0.93_{-0.07}^{+0.13}$ for the classical scalar field and $\Omega_{de}= 0.72_{-0.05}^{+0.04}$, $w_0=-0.97_{-0.03}^{+0.17}$ for the tachyonic one. The adiabatic sound speed $c_a^2$, which is the EoS parameter at early epoch, is essentially unconstrained: its 1$\sigma$ confidence range is wide and coincides practically with the prior range [-1, 0]. The mean likelihood and posterior are different, for both fields the likelihood is bimodal. First peak is close to -1, another one to 0. For classical scalar field the bimodal character of distribution is more appreciable. It means that the used datasets are not  appropriate for the estimation of this parameter. On the other hand it means that for any fixed value of $c_a^2$ from the range [-1, 0] there exists a set of the rest best fitting parameters for which the model predictions match the used observational data well. The results presented in the columns 1-7 of Table \ref{tabl2} illustrate that. In order to remove the uncertainties in determination of $c_a^2$ the other data and/or other statistical methods of analysis must be used, but this will be the matter of a separate paper.

 \begin{table*}
 \caption{The best fitting values and 1$\sigma$
 confidential ranges of cosmological parameters in the
 $\Lambda$CDM, CSF+CDM and TSF+CDM models determined
 on the base of complete data set listed in the
 section \ref{method_data}. The current
  Hubble parameter $H_0$ is in units km$\,$s$^{-1}\,
 $Mpc$^{-1}$, the age of the Universe $t_0$ is given 
 in Giga years.}
 \medskip
 \begin{tabular}{c|c|c|c|c|c|c|c|c|c}
 \hline
 & & & & & & & & &\\
 Parameters&$\Lambda$CDM&CSF+CDM&TSF+CDM&CSF+CDM&
 TSF+CDM&CSF+CDM&TSF+CDM&CSF+CDM&TSF+CDM \\
 & &$w$=const&$w$=const&$c_a^2=0$&$c_a^2=0$&$c_a^2>w_0$
  &$c_a^2>w_0$&$c_a^2$ free&$c_a^2$ free\\
  &1 &2 &3 &4 &5 &6 &7 &8 &9\\
 \hline
 & & & & & & & & &\\
$\Omega_{de}$&0.72$_{-0.05}^{+0.04}$&0.72$_{-0.04}^{+0.04}$&0.72$_{-0.05}^{+0.04}$&0.72$_{-0.05}^{+0.04}$&0.72$_{-0.05}^{+0.04}$&0.71$_{-0.05}^{+0.04}$&0.71$_{-0.04}^{+0.05}$&0.72$_{-0.06}^{+0.04}$&0.72$_{-0.05}^{+0.04}$\\
 & & & & & & & & &\\
$w_0$&-1&-1.00$_{-0.00}^{+0.17}$&-0.99$_{-0.01}^{+0.16}$&-0.99$_{-0.01}^{+0.03}$&-1.00$_{-0.00}^{+0.03}$&-0.99$_{-0.01}^{+0.16}$&-1.00$_{-0.00}^{+0.17}$&-0.93$_{-0.07}^{+0.13}$&-0.97$_{-0.03}^{+0.17}$\\
 & & & & & & & & &\\
$c_a^2$&-1&-1.00$_{-0.00}^{+0.17}$&-0.99$_{-0.01}^{+0.16}$&0&0&-0.05$_{-0.94}^{+0.05}$&-0.06$_{-0.94}^{+0.06}$&-0.97$_{-0.03}^{+0.97}$&-0.99$_{-0.01}^{+0.99}$\\
 & & & & & & & & &\\
$100\Omega_b h^2$&2.27$_{-0.14}^{+0.13}$&2.25$_{-0.12}^{+0.16}$&2.26$_{-0.15}^{+0.15}$&2.26$_{-0.15}^{+0.14}$&2.24$_{-0.13}^{+0.17}$&2.25$_{-0.12}^{+0.17}$&2.25$_{-0.13}^{+0.17}$&2.25$_{-0.13}^{+0.17}$&2.29$_{-0.16}^{+0.14}$\\
 & & & & & & & & &\\
$10\Omega_{cdm} h^2$&1.14$_{-0.08}^{+0.09}$&1.15$_{-0.11}^{+0.09}$&1.14$_{-0.11}^{+0.09}$&1.14$_{-0.10}^{+0.09}$&1.14$_{-0.13}^{+0.10}$&1.13$_{-0.13}^{+0.10}$&1.15$_{-0.14}^{+0.08}$&1.11$_{-0.12}^{+0.13}$&1.13$_{-0.13}^{+0.10}$\\
 & & & & & & & & &\\
$H_0$&70.0$_{- 3.8}^{+ 3.4}$&69.4$_{- 3.9}^{+ 4.0}$&69.7$_{- 4.4}^{+ 4.0}$&69.2$_{- 4.3}^{+ 3.8}$&69.5$_{- 3.6}^{+ 4.0}$&68.6$_{- 4.4}^{+ 4.7}$&69.2$_{- 4.7}^{+ 4.2}$&69.2$_{- 5.1}^{+ 4.2}$&69.7$_{- 5.6}^{+ 3.7}$\\
 & & & & & & & & &\\
$n_s$&0.97$_{-0.03}^{+0.03}$&0.97$_{-0.03}^{+0.04}$&0.97$_{-0.04}^{+0.03}$&0.98$_{-0.04}^{+0.05}$&0.97$_{-0.03}^{+0.03}$&0.97$_{-0.04}^{+0.05}$&0.97$_{-0.03}^{+0.04}$&0.97$_{-0.03}^{+0.05}$&0.97$_{-0.04}^{+0.04}$\\
 & & & & & & & & &\\
$\log(10^{10}A_s)$&3.09$_{-0.09}^{+0.09}$&3.09$_{-0.09}^{+0.09}$&3.09$_{-0.10}^{+0.09}$&3.09$_{-0.10}^{+0.09}$&3.09$_{-0.09}^{+0.09}$&3.09$_{-0.09}^{+0.10}$&3.09$_{-0.09}^{+0.10}$&3.07$_{-0.08}^{+0.11}$&3.09$_{-0.09}^{+0.10}$\\
 & & & & & & & & &\\
$z_{rei}$&10.5$_{- 3.3}^{+ 3.0}$&10.6$_{- 3.4}^{+ 3.2}$&10.5$_{- 3.4}^{+ 3.2}$&10.7$_{- 3.3}^{+ 3.7}$&10.5$_{- 3.5}^{+ 3.2}$&10.8$_{- 3.7}^{+ 3.2}$&10.3$_{- 3.2}^{+ 3.4}$&10.3$_{- 3.3}^{+ 3.9}$&10.4$_{- 3.2}^{+ 3.4}$\\
 & & & & & & & & &\\
$\sigma_{8}$&0.83$_{-0.06}^{+0.06}$&0.83$_{-0.10}^{+0.06}$&0.83$_{-0.10}^{+0.06}$&0.78$_{-0.19}^{+0.10}$&0.82$_{-0.12}^{+0.07}$&0.78$_{-0.18}^{+0.10}$&0.83$_{-0.19}^{+0.06}$&0.79$_{-0.19}^{+0.10}$&0.81$_{-0.20}^{+0.08}$\\
 & & & & & & & & &\\
$t_0$&13.7$_{- 0.3}^{+ 0.3}$&13.8$_{- 0.3}^{+ 0.3}$&13.7$_{- 0.3}^{+ 0.3}$&13.7$_{- 0.3}^{+ 0.4}$&13.8$_{- 0.3}^{+ 0.3}$&13.8$_{- 0.4}^{+ 0.4}$&13.8$_{- 0.3}^{+ 0.4}$&13.8$_{- 0.3}^{+ 0.4}$&13.7$_{- 0.3}^{+ 0.4}$\\
 & & & & & & & & &\\
 \hline
$-\log L$&4027.60&4027.63&4027.62&4027.24&4027.75&4027.51&4027.79&4027.35&4027.41\\
 \hline
 \end{tabular}
 \label{tabl2}
 \end{table*}

Analogical likelihood and posterior distributions for the rest of parameters have the shape of the Gaussians with half width equal approximately to the values shown in the upper/lower indices of each value in the tables. We can see also that the best fitting values of all parameters of models presented in the Table \ref{tabl2} in columns 1-7 lie in the 1$\sigma$ marginalized confidential limits of the models with free $c_a^2$. The differences of $\chi^2$ values are statistically insignificant. It means that observational datasets used here give no possibility to distinguish these 9 models at high statistical level, none of which can be ruled out by these data. 

\begin{figure*}
\includegraphics[width=.47\textwidth]{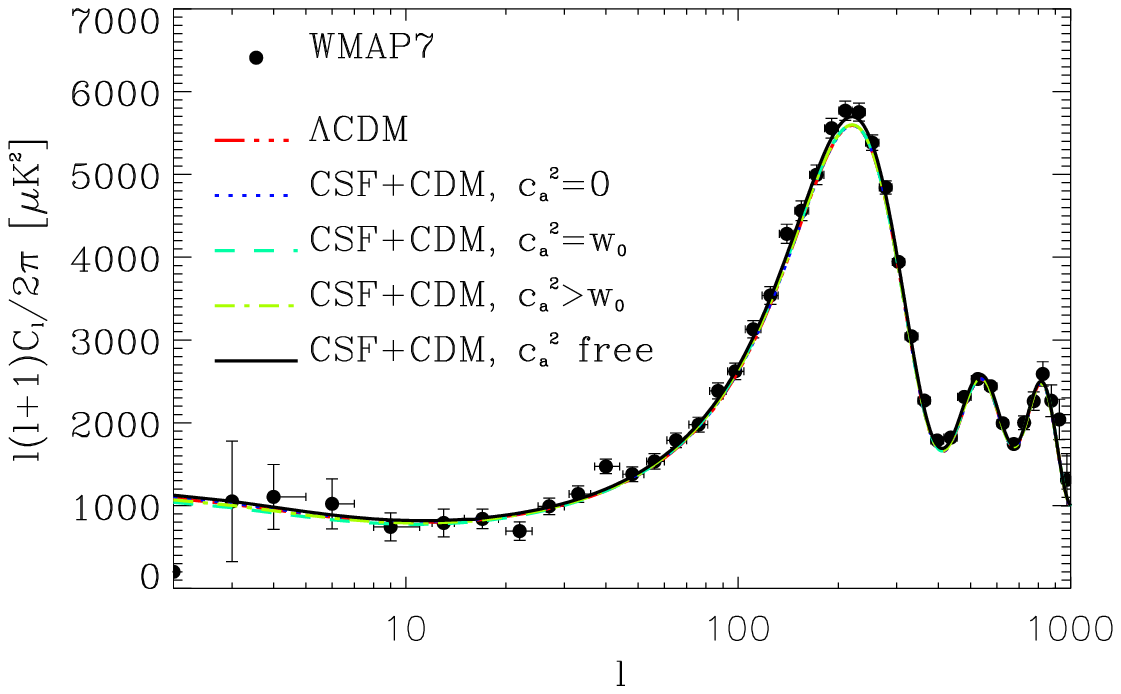}
\includegraphics[width=.47\textwidth]{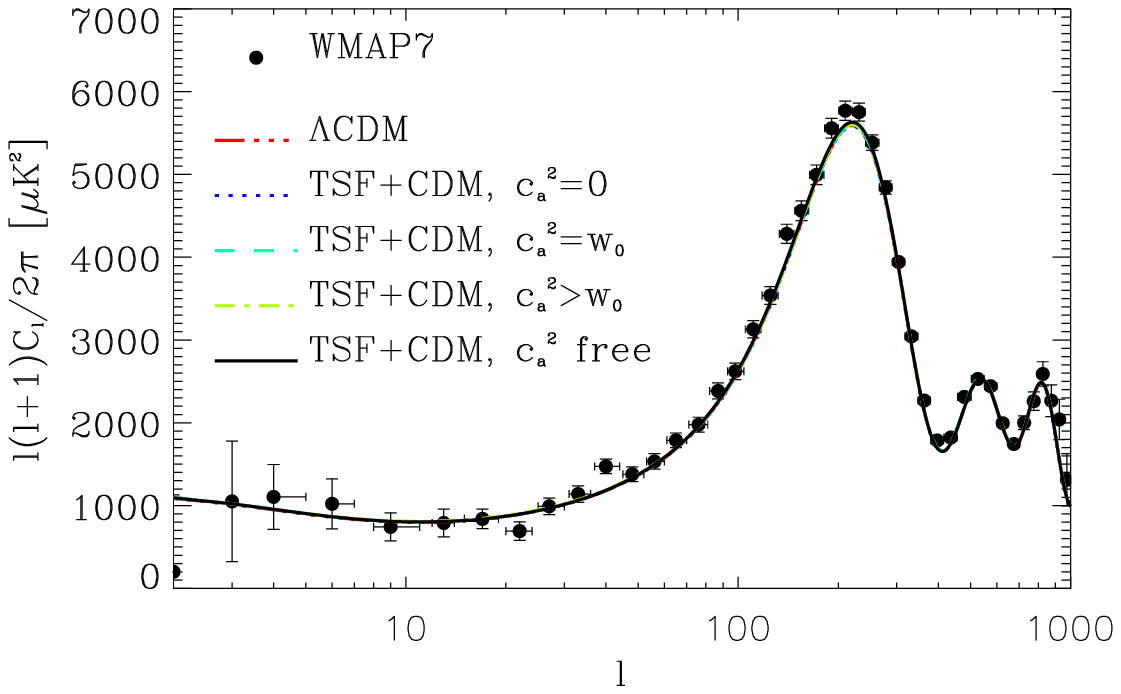}
\includegraphics[width=.47\textwidth]{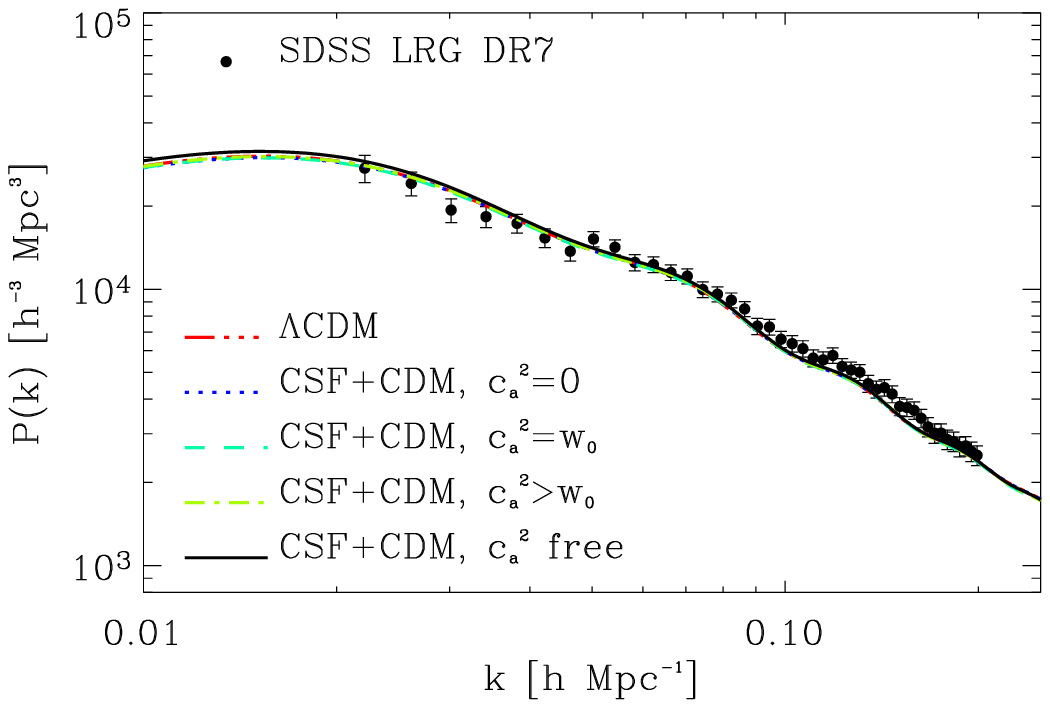}
\includegraphics[width=.47\textwidth]{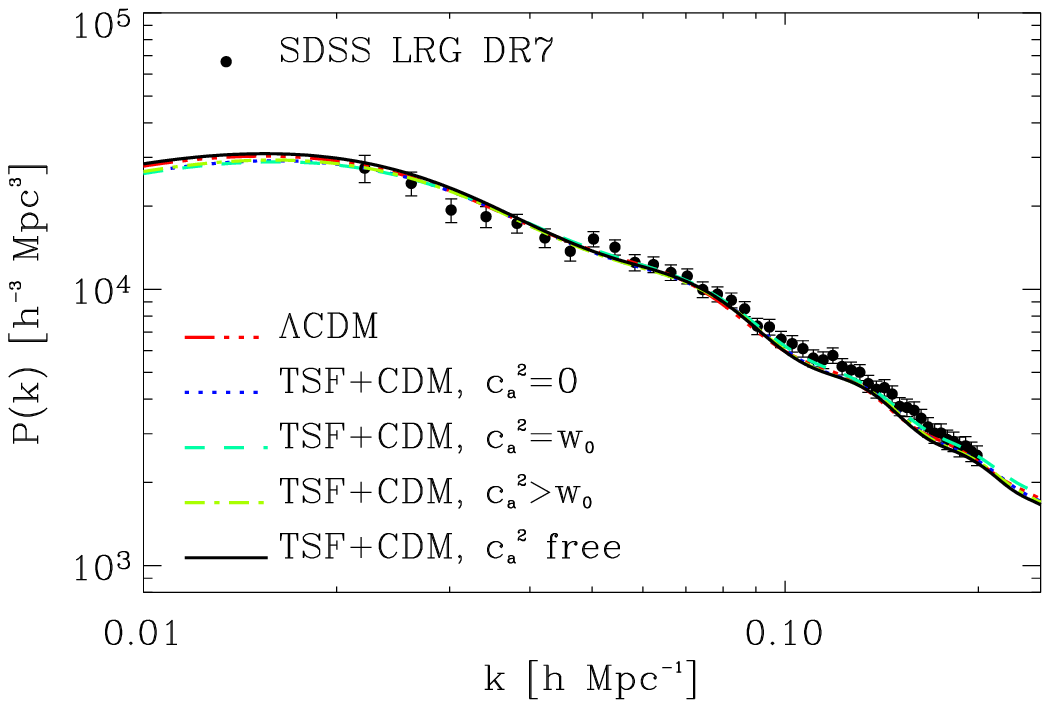}
\caption{The power spectra of CMB temperature fluctuations (top) and matter density ones (bottom) for cosmological models with classical (left) and tachyonic (right) scalar fields and best fitting parameters from Table \ref{tabl2}. The corresponding observational data from WMAP7 and SDSS LRG DR7 are presented.}
\label{cl_pk}
\end{figure*}

In Fig. \ref{cl_pk} we compare the power spectra of CMB temperature fluctuations and matter density ones for cosmological models with the best fitting parameters from Table \ref{tabl2} with WMAP7 and SDSS DR7 LRG data. They demonstrate perfect agreement of all models with observations. The power spectra of best fitting $\Lambda$CDM model (its parameters are presented in the column 1 of Table \ref{tabl2}) overlap with the spectra of CSF+CDM and TSF+CDM models. 

Let us discuss the best fitting CSF+CDM and TSF+CDM models presented in the Table \ref{tabl2}. For comparison we present the $\Lambda$CDM model with the best fitting parameters determined for the same dataset by original CosmoMC.

1. Scalar field models with $c_a^2=0$ (columns 4 and 5 in Table \ref{tabl2}). Such fields have partially the properties of $\Lambda$CDM model. Their pressure is constant during the whole history of the Universe: $P_{de}=w_0\rho_{de}^{(0)}=const$. But their density $\rho_{de}=\rho_{de}^{(0)}[(1+w_0)a^{-3}-w_0]$ at the early epoch changes similarly to one of the dust matter ($\propto a^{-3}$) and asymptotically approaches the constant value $-w_0\rho_{de}^{(0)}$ in far future ($a\gg 1$). The EoS parameter $w(a)=w_0a^3/(1+w_0-w_0a^3)$ decreases from $\approx0$ at the early epoch ($a\approx0$) to -1 in future. Now it is very close to its asymptotic value: $w_0\approx -0.99$. The expansion rate (\ref{H}) decreases from $H(a)=\sqrt{\Omega_r}H_0a^{-2}$ in early epoch to the constant value $\mathcal{H}=\sqrt{-w_0\Omega_{de}}H_0\approx 0.85H_0$ in future. The acceleration parameter varies slowly from +1 in the early Universe to $q_0=-0.57$ for classical field and $q_0=-0.58$ for tachyonic one at the current epoch up to -1 in far future. So, such scalar fields in the early Universe mimic the dust matter and will mimic vacuum energy density in future. The final stage of evolution of the Universe with classical or tachyonic scalar field with $c_a^2=0$ will be
de Sitter one: $a\propto \exp{\mathcal{H}t}$ forever. These fields jointly with cold dark matter can be considered as some type of the generalized dark component which decays and has properties of the two-component model, dark matter plus dark energy. The evolution of density perturbations of matter and both scalar fields is shown for the wave number $k=0.1$ Mpc$^{-1}$ in Fig. \ref{ddeb} (for other scales in the two-component model see \cite{Sergijenko2009b}). The density perturbations of classical and tachyonic scalar fields evolve differently because of difference of their effective sound speeds ($c_s^2=1$ for classical field and $c_s^2=-w(a)$ for the tachyonic one). Their impact on the matter density perturbations in the models with the same parameters is distinguishable too (see Fig. \ref{de2l}). However, small variation of the main cosmological parameters (columns 3 and 4 in the Table \ref{tabl2}) reduce this distinction (Fig. \ref{cl_pk}) at the cost of a bit larger $\chi^2$ for the tachyonic field. At the decoupling ($z_{dec}\approx 10^3$) and at the earlier epochs the ratio of scalar field density to matter one was constant ($\rho_{de}/\rho_{m}\approx (1+w_0)\Omega_{de}/\Omega_m$) and for parameters from the table it equals $\approx 0.015$. So, scalar fields with such parameters do not change practically the predictions of the concordance $\Lambda$CDM model for the Big Bang nucleosynthesis, however, they reduce essentially the fine tuning problem: at the end of phase transitions ($t\sim 10^{-10}$ s) their energy densities were about $10^{-12}$ of radiation energy density against $10^{-54}$ for $\Lambda$CDM-model.

2. Scalar field models with $c_a^2>w_0$ (columns 6 and 7 in Table \ref{tabl2}). The condition $c_a^2>w_0$ changes slightly the distributions shown in the bottom row of Fig. \ref{likestats} but causes finding of the best fitting value of $c_a^2$ near 0, therefore the properties of such dark energy models are similar to ones with fixed $c_a^2=0$. The future of the Universe with such fields is de Sitter expansion $a\propto \exp{\mathcal{H}t}$, where $\mathcal{H}=\sqrt{\Omega_{de}(c_a^2-w_0)/(1+c_a^2)}H_0\approx 0.84H_0$ for both fields.

3. Scalar field models with $w=const$ (columns 2 and 3 in Table \ref{tabl2}). The density of scalar fields with the best fitting parameters decreases very slowly ($\rho_{de}\propto a^{-0.0003}$ for classical field and  $\propto a^{-0.03}$ for tachyonic one) during the whole history of the Universe, the pressure  follows it rigorously: $P_{de}\approx-\rho_{de}$. The ratio of scalar field density to matter one was $\rho_{de}/\rho_{m}\approx 10^{-8}$ at the decoupling and lower at the earlier epoch. So, such fields also do not change the expansion dynamics of the early Universe and predictions of the concordance $\Lambda$CDM model. In the very early Universe, at $t\sim 10^{-10}$ s, the energy densities of classical and tachyonic scalar fields with the best fitting parameters were about $10^{-54}$ of radiation one, so, the models with such fields have the same fine tuning problem. The acceleration parameters at current epoch are the same for both fields: $q_0=-0.57$. Therefore, the past and future dynamics of expansion of the Universe with such fields is similar to the $\Lambda$CDM model one. The best fitting parameters of CSF+CDM and TSF+CDM models with $w=const$ are close, since the evolution of their density perturbations is similar for values of $w$ relatively close to -1 (see Fig. \ref{ddeb} and  \ref{de2l}). So, such fields with different Lagrangians are practically indistinguishable by cosmological observations.  

4. Scalar field models with $-1<c_a^2<w_0$ (columns 8 and 9 in Table \ref{tabl2}). The best fitting values of $c_a^2$ and $w_0$ are -0.97 and -0.93 for classical scalar field, -0.99 and -0.97 for tachyonic one correspondingly. The EoS parameter $w$ increases very slowly from the first value to the second one during the age of the Universe. In the early epoch the density of classical scalar field  changed $\propto a^{-0.09}$, the density of tachyonic one $\propto a^{-0.04}$. Their energy densities were only $\sim 10^{-51}$ of radiation one at the baryogenesis epoch ($t\sim 10^{-10}$) and $\sim 10^{-31}$ at the nucleosynthesis epoch. At the decoupling they were $\sim 10^{-8}$ of matter density. For the $\Lambda$CDM model these values are $\sim 10^{-54}$, $\sim 10^{-32}$ and $\sim 10^{-8}$ respectively. So, in this case -- like the previous one -- the classical and tachyonic scalar fields do not change the expansion dynamics of the Universe in its past history and do not affect the local physical processes such as cosmological recombination, nucleosynthesis, baryosynthesis etc. comparing to the concordance $\Lambda$CDM model. The current acceleration parameters are $q_0=-0.51$ in the model with classical scalar field and $q_0=-0.55$ in the model with tachyonic one. But predicted by these models future dynamics of the Universe differs essentially from the one predicted by previous models. The general properties were discussed in section \ref{bckgr}, here we particularize them for the best fitting parameters of the model with classical scalar field. So, in this model the EoS parameter will continue to increase in future. It will reach -1/3 at $a_{(q=0)}\approx 530$, when accelerated expansion of the Universe is changed by the decelerated one (the age of the Universe will be $\approx 195$ Gyrs). Thereafter $w$ will reach 0 at $a_{(w=0)}\approx 678$ when the age of the Universe is $\approx 208$ Gyrs. If the scalar field is tachyonic then at this moment its potential will become imaginary. The potential of classical scalar field will become negative when $w$ reaches 1 at $a_{w=1}\approx 804$, the age of the Universe will be $\approx 226$ Gyrs then. The next important moment in the evolution of the scalar field with such parameters will occur when the Universe is $\approx 268$ Gyrs old ($a=950$): it will violate the weak energy condition ($\rho_{de}\ge0$) and its energy density will become negative after that. The pressure at this time will have constant positive value ($P_{de}\approx(w_0-c_a^2)\approx 0.03\rho_{de}^{(0)}$), EoS parameter will have a discontinuity of the second kind (like in Fig. \ref{fig1}), as well as the potentials of both fields and kinetic terms of tachyonic scalar field (like in Fig. \ref{U_X}). Some time after $a_{(\rho_{de}=0)}$ the total energy density remains positive ($\rho_m+\rho_{de}>0$), the strong energy condition $\rho_m+\rho_{de}+3P_{de}>0$ is still satisfied. The decelerated expansion of the Universe will stop finally when the scale factor reaches the maximal value $a_{max}$ which is slightly larger than $a_{(\rho_{de}=0)}$. At this moment $\dot{a}=0$ and $\ddot{a}<0$. The values of matter and dark energy densities at the turnaround are $\rho_m=-\rho_{de}\approx 3\cdot 10^{-10}\rho^{(0)}_{de}$. The age of the Universe will be $\approx 268$ Gyrs then. After that the collapse begins. So, the Universe, filled with the scalar field with such parameters, is limited in time by the age of $\approx 540$ Gyrs and finishes its existence in the Big Crunch singularity. Of course, the possibility of existence of scalar fields with such unusual properties needs yet the comprehensive analysis. 

Therefore, all models presented in Table \ref{tabl2} match well the set of current observational data used here (Fig. \ref{cl_pk}). Using different subsets (for example, excluding SDSS LRG DR7, SN data or both) or substituting the last release of SDSS LRG DR7 by one of the previous ones, e. g. by DR4, changes somewhat the best fitting values,
so the preferable model has $c_a^2>w_0$. The same relation between these parameters holds when the SDSS date are excluded as well as WMAP7 and Union2 SN data are substituted by the WMAP5 \cite{Komatsu2009} and Union \cite{SNUnion} datasets correspondingly (see also \cite{Apunevych2010}). It means that the different datasets, obtained by different techniques and corresponding to different redshifts, are not in complete agreement with each other yet.

\section{Conclusion}\label{conclusions}

We have analyzed the minimally coupled cosmological scalar fields with the time-variable EoS parameter $w(a)$ and constant adiabatic sound speed $c_a^2$, which means that the temporal derivative of pressure follows rigorously the temporal derivative of energy density: $\dot{P_{de}}\propto \dot\rho_{de}$. In this case the equation of state has two natural parameters, $w_0$ and $c_a^2$, which are the EoS parameter at the present epoch and at the early stages correspondingly. The evolution of $w(a)$ from $c_a^2$ at the beginning to $w_0$ now is monotonous, however the future character of it is determined
by the relation between the values of $w_0$ and $c_a^2$. So, in the case $c_a^2>w_0$ the derivative of $w$ with respect to the scale factor $a$ is negative, $w'<0$, and fields roll down to the vacuum energy at the infinite time like one with $w=-1$. In the opposite case $c_a^2<w_0$, $w'>0$ and fields lose their repulsion properties. The accelerated expansion of the Universe will change in future to the decelerated one, the weak energy condition will be violated and the Universe will eventually collapse. Intermediate case $c_a^2=w_0$ is the well studied simplest dark energy model with $w=const$. The dynamics of expansion of the multicomponent Universe in the past depends slightly on the value of $c_a^2$ (see Fig. \ref{fig2}) but depends strongly on the values of $\Omega_{de}$ and $w_0$ related to the current epoch. This gives the possibility to constrain them by observational data. In this paper we have concentrated on determination of all parameters of scalar field models of dark energy. For this purpose we have analyzed the evolution of density and velocity perturbations of scalar fields (equations (\ref{d_de}) and (\ref{V_de})) and their influence on the evolution of matter ones. The entropy perturbations are inherent for scalar fields, so the dark energy pressure perturbations are connected with the density ones via the effective sound speed $c_s^2$ instead of the adiabatic one $c_a^2$. For the classical and tachyonic Lagrangians it equals 1 and $-w$ correspondingly. The results, presented in Fig. \ref{ddeb} and \ref{de2l}, show that the fields with different $c_a^2$ and $c_s^2$ are distinguishable in principle if the rest of cosmological parameters is known. Unfortunately, it is not the case and we have to determine the parameters of scalar field models of dark energy $\Omega_{de}$, $w_0$, $c_a^2$ jointly with other cosmological parameters, the minimal set of which is: $\Omega_b$, $\Omega_{cdm}$, $n_s$, $A_s$, $H_0$, $\tau$, $A_{SZ}$. We restricted ourselves to the spatially flat models -- that reduces the number of free model parameters to 9.

For the calculations of evolution of perturbations in all components, the power spectra of matter density perturbations and cosmic microwave background anisotropy we have used the publicly available code CAMB \cite{camb,camb_source}, modified to include the presented here expressions (\ref{w})- (\ref{H}) and (\ref{d_de})-(\ref{v_de_init}). To determine the parameters of scalar field models of dark energy jointly with rest of cosmological ones we have performed the MCMC analysis for set of current observational data, which include the power spectra from WMAP7 \cite{wmap7} and SDSS DR7 \cite{Reid2009}, the light curves of  SNIa  \cite{SNUnion2}, the Hubble constant measurements \cite{Riess2009} and  Big Bang nucleosynthesis (BBN) prior \cite{bbn}.  
It has been obtained that the best fitting parameters of scalar field models of dark energy have next values and 1$\sigma$ confidence limits: $\Omega_{de}=0.72_{-0.06}^{+0.04}$, $w_0=-0.93_{-0.07}^{+0.13}$, $c_a^2=-0.97_{-0.03}^{+0.97}$ in the case of Klein-Gordon Lagrangian and $\Omega_{de}=0.72_{-0.05}^{+0.04}$, $w_0=-0.97_{-0.03}^{+0.17}$, $c_a^2=-0.99_{-0.01}^{+0.99}$ in the case of Dirac-Born-Infeld one. The first two parameters, corresponding to the current epoch, are determined well, while the last one much worse (Fig. \ref{likestats}). So the used observational data prefer the  scalar fields with $w'>0$, which lose their repulsive properties, and predict the collapse of spatially flat Universe. But other models with $w'\le0$, $c_a^2\ge w_0$ and even $c_a^2=0$, are in the $1\sigma$ range of the best fitting one (Table \ref{tabl2}). Therefore, the values of EoS parameters of the scalar fields at the early epoch are not determined surely enough to say which field dominates now, with receding ($w'>0$) or raising ($w'<0$) repulsion, and, correspondingly, which is the future fate of our Universe, eternal accelerated expansion or decelerated one and collapse. We also conclude that the used observational data give no possibility to distinguish two different scalar fields (Fig. \ref{cl_pk} and Table \ref{tabl2}), classical and tachyonic, but we hope that the operating and planed observational programs will give the possibility to constrain better the parameters of scalar field models of dark energy and the number of admissible evolution trajectories of our Universe.

\begin{acknowledgments}
This work was supported by the project of Ministry of Education and Science of Ukraine (state registration number 0110U001385), research program ``Cosmomicrophysics'' of the National Academy of
Sciences of Ukraine (state registration number 0109U003207) and the SCOPES project No. IZ73Z0128040 of Swiss National Science Foundation. Authors also acknowledge the usage of CAMB and CosmoMC packages and are thankful to Main Astronomical Observatory of NASU for the possibility to use
the computer cluster for MCMC runs.

\end{acknowledgments}

\end{document}